\documentclass[a4paper,10pt,aps,reprint,prl,nofootinbib,showpacs,noshowkeys]{revtex4-1}
\usepackage{hyperref} 
\usepackage{amsmath}
\usepackage{amsfonts}
\usepackage{amsthm}
\usepackage{amssymb}
\usepackage{cancel}
\usepackage{graphicx}
\usepackage[toc,page]{appendix}
\usepackage{bbding}
\usepackage{slashed}
\usepackage[sort&compress]{natbib}
\usepackage{tikz} % for cluster graphs

\newcommand*\Action{\tilde{S}} % action
\begin{document}

\title{Magnetic monopole mass bounds from heavy ion collisions and neutron stars}
\date{April 26, 2017}
\author{Oliver Gould}
\email{o.gould13@imperial.ac.uk}
\author{Arttu Rajantie}
\email{a.rajantie@imperial.ac.uk}
\affiliation{Department of Physics, Imperial College London, SW7 2AZ, UK}


\begin{abstract}
Magnetic monopoles, if they exist, would be produced amply in strong magnetic fields and high temperatures via the thermal Schwinger process. Such circumstances arise in heavy ion collisions and in neutron stars, both of which imply lower bounds on the mass of possible magnetic monopoles. In showing this, we construct the cross section for pair production of magnetic monopoles in heavy ion collisions, which indicates that they are particularly promising for experimental searches such as MoEDAL.
\end{abstract}

\maketitle

There are compelling theoretical reasons to expect the existence of magnetic monopoles \cite{dirac1931quantised,thooft1974magnetic,*polyakov1974particle,polchinski2003monopoles} and, as a consequence, there have been extensive searches for them \cite{patrizii2015status,rajantie2016search}, but so far with no positive results. Astrophysical \cite{parker1970origin,*turner1982magnetic,*adams1993extension} and cosmic ray \cite{ambrosio2002final,*detrixhe2010ultrarelativistic,*adrianmartinez2011search,*aartsen2015searches,*burdin2014noncollider} searches have provided constraints on the monopole flux in the Universe, and collider searches \cite{fairbairn2006stable,aad2015search,olive2016review,acharya2016search,englert2016getting} have constrained the production cross section over a given mass range. However, in the absence of reliable theoretical predictions for the flux or the cross section, these cannot be converted into direct bounds on the monopole mass.

In collider searches, the tree-level Drell-Yan cross section has often been used to obtain indicative mass constraints \cite{aad2015search,olive2016review,acharya2016search}. However, this is not a reliable estimate because, if magnetic monopoles exist, they are strongly coupled. The Dirac quantisation condition implies that the minimum magnetic charge is given by $g_D:=2\pi/e$, where $e$ is the charge of the positron and we use natural units, $c=\hbar=k_B=\epsilon_0=1$. The magnetic fine structure constant is then $\alpha_M\approx 34$. As a result, there have been no rigorous calculations of any cross sections for magnetic monopole pair production.

In low entropy collisions of particles it has been argued that the pair production of 't Hooft-Polyakov magnetic monopoles \cite{thooft1974magnetic,*polyakov1974particle} is exponentially suppressed by \cite{drukier1982monopole} 
\begin{equation}
 \mathrm{e}^{-16\pi/e^2}\approx 10^{-236}, \label{eq:drukier_suppression}
\end{equation}
even at arbitrarily high energies. If this suppression is indeed present, it would effectively rule out the production of composite magnetic monopoles in, for example, p-p collisions at the LHC. For elementary (Dirac) monopoles \cite{dirac1948theory,*cabibbo1962quantum,*schwinger1966magnetic,*zwanziger1970local,*blagojevic1985quantum,*milton2006theoretical} the arguments of Ref. \cite{drukier1982monopole} do not apply and cross sections for pair production are completely unknown.

In this letter we consider magnetic monopole pair production in strong magnetic fields and high temperatures. We use the results of Ref. \cite{gould2017thermal}, due to the present authors, where the rate of thermal Schwinger pair production was calculated at arbitrary coupling, generalising an earlier calculation at zero temperature \cite{affleck1981monopole,*affleck1981pair}. From this, we derive an expression for the cross section of magnetic monopole pair production in heavy ion collisions. For high enough collision energies, the result is not exponentially suppressed as in Eq. \eqref{eq:drukier_suppression}. We believe that this is because the energy is spread across many degrees of freedom in the initial thermal state. This is what was found in the case of $(B+L)$ violation \cite{thooft1976symmetry,*thooft1976computation,klinkhamer1984saddle,*kuzmin1985anomalous,*arnold1987sphalerons,*arnold1987sphaleron}, in the language of which the process we consider is a sphaleron induced decay. By comparison to 
an experimental upper bound on the cross section \cite{he1997search}, we derive lower bounds on the mass of possible magnetic monopoles.  These bounds are model-independent in the sense that they apply to both elementary and composite (e.g. 't Hooft-Polyakov) monopoles and do not rely on (inapplicable) perturbation theory.

We also consider magnetic monopole pair production in the strong, long-lived magnetic fields present around neutron stars. Sufficiently light magnetic monopoles would be produced by thermal Schwinger pair production and dissipate the magnetic field. By comparison with the observed magnetic field strengths we derive another set of lower mass bounds.

For comparison, the current best, model-independent, lower bound on the mass of magnetic monopoles derives from reheating and big bang nucleosynthesis (BBN). Inflation would have diluted away any pre-existing magnetic monopoles \cite{zeldovich1978concentration,*guth1979phase,*preskill1979cosmological} but,  during reheating, sufficiently light magnetic monopoles would have been produced thermally. Hence, from the bounds on the monopole flux in the Universe \cite{parker1970origin,*turner1982magnetic,*adams1993extension,ambrosio2002final,*detrixhe2010ultrarelativistic,*adrianmartinez2011search,*aartsen2015searches,*burdin2014noncollider}, one can derive a bound on the ratio of the mass of any magnetic monopoles to the reheating temperature, $m/T_{\mathrm{RH}}\gtrsim 45$ \cite{turner1982thermal,*collins1984thermal,*lindblom1984thermal}. Further, as the reheating temperature must be greater than the temperature of BBN, $T_{\mathrm{BBN}}\approx 10\mathrm{MeV}$, we find that the mass of any magnetic monopoles must 
satisfy $m\gtrsim 0.45\mathrm{GeV}$.

Magnetic monopoles are strongly coupled to the photon field and hence the usual weak coupling results for Schwinger pair production are inapplicable. However, when the external field is weak and slowly varying, the calculation of the rate of Schwinger pair production becomes semiclassical irrespective of the magnitude of the coupling. In particular the small semiclassical parameter, akin to $\hbar$, is $gB/m^2$, where $g$ is the magnetic coupling, $B$ is the external magnetic field and $m$ is the mass of the magnetic monopoles.  

In the leading semiclassical approximation, the rate of pair production per unit volume, $\Gamma_T$, is of the form
\begin{equation}
 \log(\Gamma_T) = -\frac{m^2}{gB}\bigg\{\Action\left(g,m,B,T\right) + O\left(\frac{gB}{m^2}\log\left(\frac{gB}{m^2}\right)\right)\bigg\}.
\end{equation}
where $m^2\gg gB$ and the action, $\Action$, is a function only of the dimensionless ratios $g^3B/m^2$ and $mT/gB$. It is not smooth, having discontinuities which can be described as phase transitions. It has been calculated in Ref. \cite{gould2017thermal}, analytically in various limits as well as numerically.

Within these approximations, of a weak and slowly varying external field, the results are not model specific. They apply to both elementary and composite monopoles, whether scalar or spinor (see appendices A and B of Ref. \cite{gould2017thermal}). This is because, in the physical regime we consider, any structure of the magnetic monopoles is invisible, as in Refs. \cite{boulware1976scattering,*bardakci1978local,*manton1978effective}. For composite monopoles we must also assume that the monopoles are small compared with other scales in the problem. For the usual grand unified theory monopoles, this approximation fails when $m^2 \lesssim g^3B/4\pi$. For elementary monopoles, virtual monopole pairs will modify the photon-monopole interaction on sufficiently short scales. We can make a simple estimate of the scale at which this effect becomes significant by setting the rest mass of a monopole pair equal to the Coulomb attraction. This defines the scale $r\sim g^2/8\pi m$. The semiclassical calculation of 
Ref. \cite{gould2017thermal}, which does not include this effect, thus breaks down when the scale of the instanton probes these short length scales, i.e. when $m^2\lesssim g^3 B/8\pi$.

In this paper, we will be interested in two particular cases. For heavy ion collisions, the relevant temperatures are high. When $m^2 \gtrsim g^3B$, high temperatures are such that $T\gtrsim \sqrt{2}\pi^{-3/4}(gB^3/m^2)^{1/4}$. In this regime the action is given by
\begin{equation}
\Action\left(g,m,B,T\right)=2 \bigg( 1 - \sqrt{\frac{g^3B}{4\pi m^2}}\bigg) \frac{gB}{mT}. \label{eq:action_straight}
\end{equation}
When $g^3B/m^2$ is larger, it may be that the action is smaller than that given by Eq. \eqref{eq:action_straight}. This depends on the nature of the phase diagram as discussed in Ref. \cite{gould2017thermal}. However, the action cannot be larger than that given by Eq. \eqref{eq:action_straight} and hence the rate of pair production cannot be lower.

For neutron stars the relevant temperatures are low, $T\ll gB/m$. In this case the action is given by
\begin{align}
   \Action\left(g,m,B,T\right)& = \pi-\frac{g^3 B}{4m^2}-\zeta(4)\frac{g^3 B}{m^2}\left(\frac{mT}{gB}\right)^4\nonumber \\
   -4& \zeta(6)  \frac{g^3 B}{m^2}\left(\frac{mT}{gB}\right)^6+O\left(\frac{mT}{gB}\right)^8. \label{eq:low_temp_leading2}
\end{align}
At zero temperature, and at leading order in $g^3B/m^2$, the prefactor of the rate (as in $A$ in Ref. \cite{callan1977fate}) has been calculated \cite{affleck1981monopole,affleck1981pair}. Together they give,
\begin{equation}
 \Gamma_0 = (2s+1)\frac{g^2 B^2}{8\pi^3}\ \mathrm{e}^{-\frac{\pi m^2}{gB}+\frac{g^2}{4}}\left(1+O\left(\frac{g^3B}{m^2}\right)\right),\label{eq:gamma_zerotemp}
\end{equation}
where $s$ is the spin of the charged particle.

In a high energy heavy ion collision a fireball is created which thermalises quickly and within which there are strong magnetic fields because of the fast-moving electrically charged nuclei \cite{kharzeev2007effects,*skokov2009estimate,*voronyuk2011electromagnetic,*bzdak2011event,*deng2012event,*mclerran2013comments}. The presence of both the thermal bath and the magnetic fields means that thermal Schwinger pair production of magnetic monopoles is possible. However, only sufficiently light magnetic monopoles will be produced in measurable quantities.

For a given event, with impact parameter $b$, the fireball will be contained in some spacetime region, $\mathcal{R}(b)$. If the temperature, $T(x;b)$, and magnetic field, $B(x;b)$ are sufficiently slowly varying, then to find the total probability, $p(b)$, that a pair of magnetic monopoles is produced in a given collision, we can simply integrate the rate over the spacetime volume of the fireball,
\begin{equation}
p(b) = \int_{\mathcal{R}(b)} \mathrm{d}^4 x \ \Gamma_T(m,g,B(x;b),T(x;b)).
\end{equation}
From this we can write down the cross section for pair production,
\begin{equation}
 \sigma_{M\bar{M}} = \int \mathrm{d}b \ \frac{\mathrm{d}\sigma^{\mathrm{inel}}_{HI}}{\mathrm{d}b} \ p(b),
\end{equation}
where $\mathrm{d}\sigma^{\mathrm{inel}}_{HI}/\mathrm{d}b$ is the total, differential, inelastic cross section for the relevant heavy ion collision. Due to the exponential dependence of $\Gamma_T$ on the magnetic field and temperature, all of these integrals can be carried out in the stationary phase approximation. However, as we have only calculated the logarithm of the rate to leading order in $gB/m^2$, we will instead make the following simple estimate
\begin{equation}
 \sigma_{M\bar{M}} \approx  \sigma^{\mathrm{inel}}_{HI}\ \mathcal{V} \ \Gamma_T(m,g,B,T), \label{eq:approx_cross_section}
\end{equation}
where $\mathcal{V}$ is the spacetime volume of a typical collision and $B$ and $T$ are taken to be the maximum values of the functions $B(x;b)$ and $T(x;b)$ respectively. This expression should capture the approximate order of magnitude of the result.

In heavy ion collisions there have been both direct searches for magnetic monopoles \cite{he1997search} and (preliminary) searches for trapped monopoles in obsolete parts of the beam pipe \cite{deroeck2012development,*joergensen2012searching}. Ref. \cite{he1997search} reported the results of a search at SPS for magnetic monopoles in fixed-target lead ion collisions with beam energy $160A\mathrm{GeV}$. In this, they derived an upper bound on the magnetic monopole pair production cross section, $\sigma_{M\bar{M}}<\sigma_{M\bar{M}}^{UB}=1.9\mathrm{nb}$. By comparing this with Eq. \eqref{eq:approx_cross_section}, we can derive a lower bound on the mass of any possible magnetic monopoles.

Assuming that the prefactor of $\Gamma_T$ multiplied by $\mathcal{V}$ is not exponentially large in $m^2/gB$, we arrive at
\begin{equation}
 \log\left(\frac{\sigma^{\mathrm{inel}}_{HI}}{\sigma_{M\bar{M}}^{UB}}\right)\lesssim  \frac{m^2}{gB}\Action\left(g,m,B,T\right). \label{eq:log_cross_section}
\end{equation}
The magnetic field strength in lead ion collisions at these energies was estimated to be $B_{160\mathrm{GeV}} \approx 0.0097\mathrm{GeV}^2$ \cite{skokov2009estimate}. From an analysis of the spectrum of neutral pions, the temperature was estimated to be $T_{160\mathrm{GeV}}\approx 0.185\mathrm{GeV}$ \cite{schlagheck1999thermalization}. We take $\sigma^{\mathrm{inel}}_{HI}\approx 6.3b$, the minimum-bias cross section for the experiment \cite{aggarwal2000observation}.

Substituting Eq. \eqref{eq:action_straight} and the parameters into Eq. \eqref{eq:log_cross_section} leads to the following bound on the mass of any magnetic monopoles
\begin{equation}
 m\gtrsim \left(2.0+2.6\left(\frac{g}{g_D}\right)^{3/2}\right)\mathrm{GeV}. \label{eq:mass_bound}
\end{equation}
Note that the experiment was only sensitive to magnetic charges $g\geq 2g_D$.
 
The semiclassical approximation, made in deriving Eq. \eqref{eq:mass_bound}, requires that the exponential suppression be large. At the lower bound this amounts to $22\gg 1$. The approximation of constant magnetic field requires that the magnetic field varies significantly on time and length scales much larger than those of the instanton. The instanton has a spatial extent of $\sqrt{g/4\pi B}\approx 18\mathrm{GeV}^{-1}$ for $g=2g_D$ in the direction of the magnetic field (transverse to the beam) and a temporal extent of $1/T\approx 5.4\mathrm{GeV}^{-1}$. At SPS energies the magnetic field varies significantly over the length and timescales of the fireball. The transverse size of the fireball is of the order of the size of a lead nucleus, $2R_{Pb}\approx 100\mathrm{GeV}^{-1}$, which is somewhat larger than the spatial size of the instanton. The lifetime of the magnetic field, $t_B\approx 2R_{Pb}/\gamma\approx 11\mathrm{GeV}^{-1}$, is reduced by, $\gamma$, the Lorentz factor in the centre of mass frame \cite{
voronyuk2011electromagnetic,*deng2012event,*mclerran2013comments} (though it has been suggested that the lifetime may be longer \cite{tuchin2010synchrotron,*tuchin2013particle,*tuchin2013time}). This lifetime, $t_B$, is somewhat larger than the temporal extent of the instanton.

At higher energies one would expect to produce higher mass magnetic monopoles, if such particles exist. The magnetic field strength increases linearly with the centre of mass energy, $\sqrt{s}$, \cite{bzdak2011event,*voronyuk2011electromagnetic,*deng2012event} and the temperature increases logarithmically \cite{hwa1985initial}, both effects increasing the range of accessible masses. However at higher energies the magnetic field becomes more transient and inhomogeneous, its lifetime and thickness along the beam axis both being proportional to $1/\sqrt{s}$. This leads to a breakdown of the constant field approximation. To account for this, the calculation of Ref. \cite{gould2017thermal} would need to be modified. One would expect the temporal variation of the magnetic field to increase the rate of pair production, and the spatial variation to decrease it \cite{brezin1970pair,*dunne2005worldline,*ilderton2015nonperturbative,*kohlfurst2015electron,*torgrimsson2017dynamically}.

There is promise for magnetic monopole searches in the next scheduled Pb-Pb collisions at the LHC in 2018, at which ALICE, ATLAS, CMS, LHCb and MoEDAL may be able to detect monopoles. The trapping detectors of MoEDAL are ideally suited for this because they have no background noise \cite{acharya2016search}. Let us make the simple, though perhaps naive, assumption that the rate derived for a constant magnetic field provides a lower bound on the true rate. Using the magnetic field, $B\approx 1.1GeV^2$, and integrated luminosity, $L_{int}\approx 4\mu b^{-1}$, from the 2015 lead ion collisions at $\sqrt{s_{NN}}=5.02\mathrm{TeV}$ \cite{massacrier2016first} and the temperature, $T\approx 0.30GeV$, and cross section,  $\sigma^{\mathrm{inel}}_{HI}\approx 7.7b$, from the lower energy collisions in 2010-2011 \cite{alice2013measurement,abelev2013centrality}, and assuming an acceptance of $O(10^{-4})$, we would predict that magnetic monopoles with masses up to approximately $(1+28(g/g_D)^{3/2})\mathrm{GeV}$ could be experimentally observed.

Note that magnetic monopole pair production in heavy ion collisions has been discussed before \cite{roberts1986dirac,*dobbins1993updated}. They also consider thermal production, though they do not include the effect of the magnetic field.

There are also strong magnetic fields and high temperatures in neutron stars. Magnetic fields have been estimated to be up to $B_{\mathrm{Magnetar}}\approx 10^{-4}\mathrm{GeV}^2$ \cite{reisenegger2003origin} for the so called magnetars. The temperatures of such neutron stars lie in the range $10^{-8}\mathrm{GeV}$ to $10^{-6}\mathrm{GeV}$ for most of the stars' lifetime, though in the early stages they can be as high as $10^{-2}\mathrm{GeV}$ \cite{pons2008magnetothermal,*potekhin2011physics}.

Magnetic monopoles present in such circumstances would be accelerated by the magnetic field thuswise dissipating its energy. A calculation of this effect can be used to put upper bounds on the number density of magnetic monopoles \cite{parker1970origin,*turner1982magnetic,*adams1993extension,harvey1985effects}. We can go a step further and equate the number density to that produced by thermal Schwinger pair production, and thuswise bound the mass of any magnetic monopoles.

The magnetic field of a neutron star can be approximated as dipolar \cite{turolla2015magnetars}. We focus on the magnetic fields above the surface of the star, which are fairly well established. We assume that on the microscopic scale $m/gB$ the magnetic field can be treated as constant. Note that, due to the superconducting core, the internal magnetic fields would be contained into flux tubes increasing the field strength locally and enhancing the production rate. Hence a consideration of the interior of the neutron star may lead to stronger bounds \cite{harvey1985effects}, though one would need to consider interactions between magnetic monopoles and matter particles \cite{ahlen1978stopping,*ahlen1982calculation,*bracci1983binding,*bracci1983interactions,*derkaoui1998energy}.
%  We do not consider the interior of the neutron star, where interactions between magnetic monopoles and matter particles would be significant \cite{ahlen1978stopping,*ahlen1982calculation,*bracci1983binding,*bracci1983interactions,*derkaoui1998energy}, nor do we consider the region within the atmosphere, which is at most a few centimetres thick \cite{potekhin2011physics}.

We consider typical neutron star mass and radius parameters, $M_{NS}=1.4 M_{\odot}$ and $R\approx 1.0\times 10^{20}\mathrm{GeV}^{-1}$ respectively. At the surface of the star, where the gravitational field is strongest, the ratio of gravitational to magnetic forces on such a magnetic monopole is
 \begin{equation}
  \frac{F_G}{F_B}\approx\frac{G_N M_{NS} m}{g B R^2} \approx 7.14\times 10^{-19}\left(\frac{g_D}{g}\right)\left(\frac{ m}{\mathrm{GeV}}\right),
 \end{equation}
where $G_N$ is Newton's constant. So, for magnetic monopoles with masses much less than $10^{19}\mathrm{GeV}$, the magnetic force dominates over the gravitational one. In this regime magnetic monopoles will be accelerated by the magnetic field over a timescale $O(m/gB)$ to nearly the speed of light and will escape both the gravitational attraction of the star and the dipolar magnetic field, leaving with a kinetic energy $O(gBR)$.
% Due to their inertia, the magnetic monopoles do not follow the dipolar magnetic field lines back round into the neutron star but follow straighter, unbound trajectories, as can be checked by integrating the dual Lorentz force law with suitable parameters.

Locally the energy density of the magnetic field and thermal bath act as a source of magnetic monopoles. If the density of magnetic monopoles is low enough, which indeed it will turn out to be, we can ignore their annihilation and hence
\begin{equation}
 \nabla_\mu n^\mu = \Gamma_T, \label{eq:number_density_divergence}
\end{equation}
where $n^\mu:=n_c u^\mu$, $n_c$ is the (comoving) number density of magnetic monopoles and $u^\mu$ is their fluid velocity. Now consider a spatial region above the surface of the neutron star, small enough so that across it the magnetic field and temperature can be treated as approximately constant but large enough so that its spatial dimensions are all large compared with the low temperature instanton size, $m/gB$. We denote the area of the surface by $A$ and the volume by $V$. Integrating Eq. \eqref{eq:number_density_divergence} over this spatial region gives 
\begin{equation}
 \frac{\mathrm{d}N}{\mathrm{d}t}\approx V\Gamma_T - f A n u,\label{eq:neutron_Ndot}
\end{equation}
where $N=n V$ is the number of magnetic monopoles in the spatial region, $n:=n^0$ is the number density measured in the frame of the neutron star, $u$ is the spatial velocity in the same frame and $f$ is a numerical coefficient of order $1$, the fraction of the surface area through which magnetic monopoles may escape. The magnetic current will be aligned with the magnetic field and $u\approx 1$.

At equilibrium, the rate of change of $N$ with time will be zero, hence the number density of magnetic monopoles is equal to
\begin{equation}
 n \approx \frac{V \Gamma_T}{f A}.\label{eq:neutron_number_density}
\end{equation}
We define by $r:=V/f A$, the coefficient in front of $\Gamma_T$, which is of the order of the radius of the spatial region. The presence of the magnetic monopoles, being accelerated by the magnetic field, will dissipate the energy of the magnetic field at a rate
\begin{equation}
\frac{\mathrm{d}}{\mathrm{d}t}\left(\frac{1}{2}B^2\right)=-\mathbf{J_M}\cdot \mathbf{B}
\end{equation}
where $\mathbf{J_M}=g n \mathbf{u}$. Using that $\mathbf{J_M}\cdot \mathbf{B}\approx g n B$ and Eq. \eqref{eq:neutron_number_density} this simplifies to
\begin{equation}
\frac{\mathrm{d}B}{\mathrm{d}t} \approx -g r \Gamma_T. \label{eq:neutron_Bdot}
\end{equation}
This dissipation will provide a ceiling for the growth of the magnetic field. Consider the \emph{fast dynamo process}, argued in \cite{thompson1993neutron} to be responsible for the strong magnetic fields in magnetars. In the presence of this process the rate of change of the magnetic field is modified to
\begin{equation}
 \frac{\mathrm{d}B}{\mathrm{d}t} \approx -g r \Gamma_T + \frac{B}{2\tau_{D}}, \label{eq:neutron_Bdot_dynamo}
\end{equation}
where $\tau_{D}$ is the characteristic enhancement time of the dynamo. For sufficiently small magnetic fields the rate, $\Gamma_T$, is strongly exponentially suppressed and the dynamo action dominates. Conversely, the exponential dependence of $\Gamma_T$ on $B$ means that $\Gamma_T$ will always dominate at sufficiently large values of $B$. In between is the point of maximum $B$, at which the two effects are equal and the right hand side of Eq. \eqref{eq:neutron_Bdot_dynamo} is zero. This argument is sound if the semiclassical approximation still holds at this point.

The rate $\Gamma_T$ is bounded below by the rate at zero temperature, Eq. \eqref{eq:gamma_zerotemp}. Thus we may use this to bound the effect of the dissipation of $B$ due to the creation of magnetic monopoles. Equating the right hand side of \eqref{eq:neutron_Bdot_dynamo} to zero, and using this zero temperature rate, we derive the following bound,
\begin{equation}
 B\lesssim \frac{\pi  m^2}{g W\left(\frac{e^{\frac{g^2}{4}} (2s+1) g^2 m^2 r \tau_D}{4 \pi ^2}\right)},
\end{equation}
where $W$ is the principal part of the Lambert-W function. Inverting the argument which led to the maximum magnetic field, we may use the observation of a strong magnetic field to give a lower bound on the mass of possible magnetic monopoles,
\begin{equation}
 m\gtrsim \sqrt{\frac{g B}{\pi}\left[\frac{g^2}{4}+\log \left(\frac{ (2s+1) g^3 B r \tau_D}{4 \pi ^3}\right)\right]}.
\end{equation}
So, the largest bounds will be found from the observation of strong magnetic fields, $B$, existing over large spatial extents, $r$, and created by processes with long characteristic times, $\tau_D$. Note though that the dependence on $s$, $r$ and $\tau_D$ is only logarithmic and hence the dependence on $B$ dominates.

\begin{figure}
 \centering
  \includegraphics[width=0.48\textwidth]{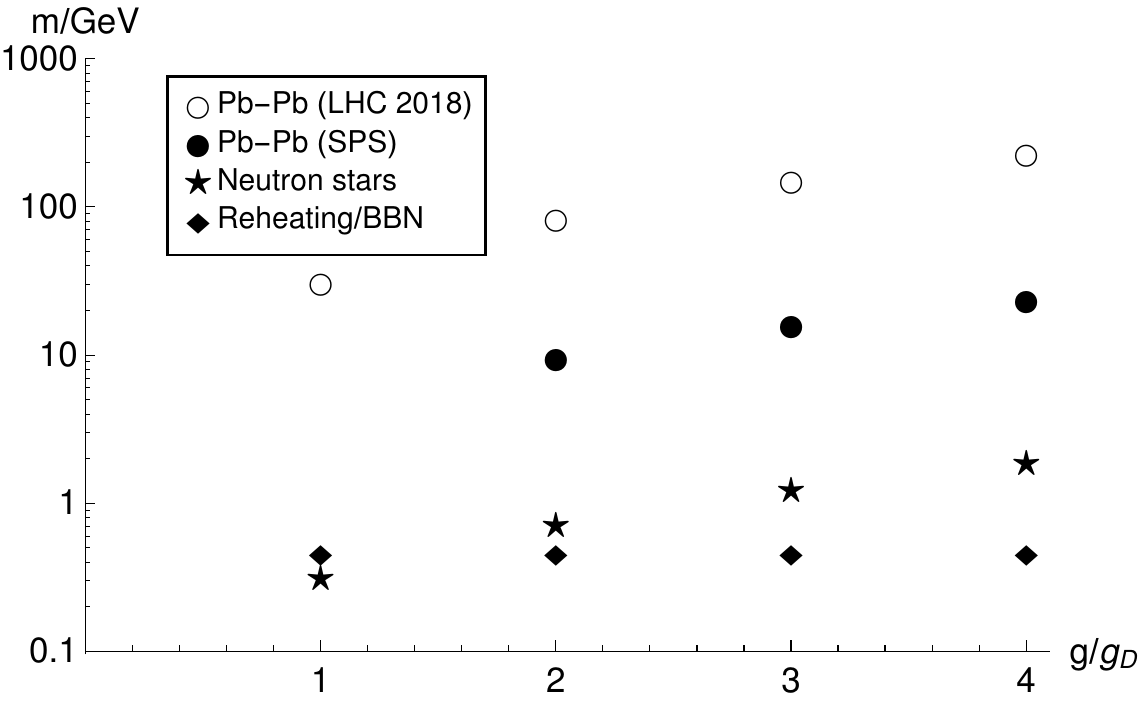}
  \caption{Summary of the lower bounds for the mass of magnetic monopoles. The empty circles are projections based on the constant field approximation, which is not expected to be accurate at LHC energies, and they should therefore be considered as tentative.}
  \label{fig:massBounds}
\end{figure}

If we take $r\approx R$, the radius of the neutron star, and $\tau_D\approx 1.5\times 10^{24}\mathrm{GeV}^{-1}$ (one second, a short characteristic dynamo time) and $B\approx B_{\mathrm{Magnetar}}$, we derive the following lower bounds: $m\gtrsim 0.31\mathrm{GeV}$ for $g=g_D$ and $m\gtrsim 0.70\mathrm{GeV}$ for $g=2g_D$. If there were to exist magnetic monopoles lighter than these lower bounds, their production and acceleration would strongly dissipate the magnetic field before it could ever reach $B_{\mathrm{Magnetar}}$. Note that for the bounding values the exponential suppression is numerically about $97\gg 1$, and hence the semiclassical approximation is valid.

A similar approach to that we have presented here was given recently in Ref. \cite{hook2017bounding}, though they considered somewhat different types of particles.

From the arguments of this letter, magnetic monopoles with masses below those indicated in Fig. \ref{fig:massBounds} cannot exist in nature. Our key approximations, that the relevant magnetic fields are weak and slowly varying, are more or less justified. Future higher energy heavy ion collisions can improve these bounds significantly though, at higher energies, accounting for the spacetime dependence of the magnetic field requires further theoretical work. In neutron stars, a consideration of the superconducting regions below the surface may lead to significant improvements in the lower mass bounds, though in this case monopole-matter interactions would have to be accounted for. The calculation of $\Gamma_T$ could be improved by computing the prefactor. For high temperatures this could first be done for $m^2\gg g^3B$, where the nonrelativistic approximation is valid.

\begin{acknowledgments}
The authors would like to thank Toby Wiseman, Sergey Sibiryakov, Ian Jubb, Edward Gillman, Lois Overvoorde, Pat Scott and Aaron Vincent for illuminating discussions. OG is supported by an STFC studentship and AR by STFC grant ST/L00044X/1.
\end{acknowledgments}

\bibliography{finiteTemperature}

\begin{thebibliography}{82}%
\makeatletter
\providecommand \@ifxundefined [1]{%
 \@ifx{#1\undefined}
}%
\providecommand \@ifnum [1]{%
 \ifnum #1\expandafter \@firstoftwo
 \else \expandafter \@secondoftwo
 \fi
}%
\providecommand \@ifx [1]{%
 \ifx #1\expandafter \@firstoftwo
 \else \expandafter \@secondoftwo
 \fi
}%
\providecommand \natexlab [1]{#1}%
\providecommand \enquote  [1]{``#1''}%
\providecommand \bibnamefont  [1]{#1}%
\providecommand \bibfnamefont [1]{#1}%
\providecommand \citenamefont [1]{#1}%
\providecommand \href@noop [0]{\@secondoftwo}%
\providecommand \href [0]{\begingroup \@sanitize@url \@href}%
\providecommand \@href[1]{\@@startlink{#1}\@@href}%
\providecommand \@@href[1]{\endgroup#1\@@endlink}%
\providecommand \@sanitize@url [0]{\catcode `\\12\catcode `\$12\catcode
  `\&12\catcode `\#12\catcode `\^12\catcode `\_12\catcode `\%12\relax}%
\providecommand \@@startlink[1]{}%
\providecommand \@@endlink[0]{}%
\providecommand \url  [0]{\begingroup\@sanitize@url \@url }%
\providecommand \@url [1]{\endgroup\@href {#1}{\urlprefix }}%
\providecommand \urlprefix  [0]{URL }%
\providecommand \Eprint [0]{\href }%
\providecommand \doibase [0]{http://dx.doi.org/}%
\providecommand \selectlanguage [0]{\@gobble}%
\providecommand \bibinfo  [0]{\@secondoftwo}%
\providecommand \bibfield  [0]{\@secondoftwo}%
\providecommand \translation [1]{[#1]}%
\providecommand \BibitemOpen [0]{}%
\providecommand \bibitemStop [0]{}%
\providecommand \bibitemNoStop [0]{.\EOS\space}%
\providecommand \EOS [0]{\spacefactor3000\relax}%
\providecommand \BibitemShut  [1]{\csname bibitem#1\endcsname}%
\let\auto@bib@innerbib\@empty
%</preamble>
\bibitem [{\citenamefont {Dirac}(1931)}]{dirac1931quantised}%
  \BibitemOpen
  \bibfield  {author} {\bibinfo {author} {\bibfnamefont {P.~A.~M.}\
  \bibnamefont {Dirac}},\ }\href {\doibase 10.1098/rspa.1931.0130} {\bibfield
  {journal} {\bibinfo  {journal} {Proc. Roy. Soc. Lond.}\ }\textbf {\bibinfo
  {volume} {A133}},\ \bibinfo {pages} {60} (\bibinfo {year}
  {1931})}\BibitemShut {NoStop}%
%\%CITATION = PRSLA,A133,60;\%\%
\bibitem [{\citenamefont {Hooft}(1974)}]{thooft1974magnetic}%
  \BibitemOpen
  \bibfield  {author} {\bibinfo {author} {\bibfnamefont {G.~t.}\ \bibnamefont
  {Hooft}},\ }\href {\doibase 10.1016/0550-3213(74)90486-6} {\bibfield
  {journal} {\bibinfo  {journal} {Nucl. Phys. B}\ }\textbf {\bibinfo {volume}
  {79}},\ \bibinfo {pages} {276} (\bibinfo {year} {1974})}\BibitemShut
  {NoStop}%
\bibitem [{\citenamefont {Polyakov}(1974)}]{polyakov1974particle}%
  \BibitemOpen
  \bibfield  {author} {\bibinfo {author} {\bibfnamefont {A.~M.}\ \bibnamefont
  {Polyakov}},\ }\href@noop {} {\bibfield  {journal} {\bibinfo  {journal} {JETP
  Lett.}\ }\textbf {\bibinfo {volume} {20}},\ \bibinfo {pages} {194} (\bibinfo
  {year} {1974})}\BibitemShut {NoStop}%
%\%CITATION = JTPLA,20,194;\%\%
\bibitem [{\citenamefont {Polchinski}(2004)}]{polchinski2003monopoles}%
  \BibitemOpen
  \bibfield  {author} {\bibinfo {author} {\bibfnamefont {J.}~\bibnamefont
  {Polchinski}},\ }\bibfield  {booktitle} {\emph {\bibinfo {booktitle}
  {{Proceedings, Dirac Centennial Symposium, Tallahassee, USA, December 6-7,
  2002}}},\ }\href {\doibase 10.1142/S0217751X0401866X} {\bibfield  {journal}
  {\bibinfo  {journal} {Int. J. Mod. Phys.}\ }\textbf {\bibinfo {volume}
  {A19S1}},\ \bibinfo {pages} {145} (\bibinfo {year} {2004})},\ \bibinfo {note}
  {[,145(2003)]},\ \Eprint {http://arxiv.org/abs/hep-th/0304042}
  {arXiv:hep-th/0304042 [hep-th]} \BibitemShut {NoStop}%
%\%CITATION = HEP-TH/0304042;\%\%
\bibitem [{\citenamefont {Patrizii}\ and\ \citenamefont
  {Spurio}(2015)}]{patrizii2015status}%
  \BibitemOpen
  \bibfield  {author} {\bibinfo {author} {\bibfnamefont {L.}~\bibnamefont
  {Patrizii}}\ and\ \bibinfo {author} {\bibfnamefont {M.}~\bibnamefont
  {Spurio}},\ }\href {\doibase 10.1146/annurev-nucl-102014-022137} {\bibfield
  {journal} {\bibinfo  {journal} {Ann. Rev. Nucl. Part. Sci.}\ }\textbf
  {\bibinfo {volume} {65}},\ \bibinfo {pages} {279} (\bibinfo {year} {2015})},\
  \Eprint {http://arxiv.org/abs/1510.07125} {arXiv:1510.07125 [hep-ex]}
  \BibitemShut {NoStop}%
%\%CITATION = ARXIV:1510.07125;\%\%
\bibitem [{\citenamefont {Rajantie}(2016)}]{rajantie2016search}%
  \BibitemOpen
  \bibfield  {author} {\bibinfo {author} {\bibfnamefont {A.}~\bibnamefont
  {Rajantie}},\ }\href {\doibase 10.1063/PT.3.3328} {\bibfield  {journal}
  {\bibinfo  {journal} {Phys. Today}\ }\textbf {\bibinfo {volume} {69}},\
  \bibinfo {pages} {40} (\bibinfo {year} {2016})}\BibitemShut {NoStop}%
%\%CITATION = PHTOA,69,40;\%\%
\bibitem [{\citenamefont {Parker}(1970)}]{parker1970origin}%
  \BibitemOpen
  \bibfield  {author} {\bibinfo {author} {\bibfnamefont {E.~N.}\ \bibnamefont
  {Parker}},\ }\href {\doibase 10.1086/150442} {\bibfield  {journal} {\bibinfo
  {journal} {Astrophys. J.}\ }\textbf {\bibinfo {volume} {160}},\ \bibinfo
  {pages} {383} (\bibinfo {year} {1970})}\BibitemShut {NoStop}%
%\%CITATION = ASJOA,160,383;\%\%
\bibitem [{\citenamefont {Turner}\ \emph {et~al.}(1982)\citenamefont {Turner},
  \citenamefont {Parker},\ and\ \citenamefont {Bogdan}}]{turner1982magnetic}%
  \BibitemOpen
  \bibfield  {author} {\bibinfo {author} {\bibfnamefont {M.~S.}\ \bibnamefont
  {Turner}}, \bibinfo {author} {\bibfnamefont {E.~N.}\ \bibnamefont {Parker}},
  \ and\ \bibinfo {author} {\bibfnamefont {T.~J.}\ \bibnamefont {Bogdan}},\
  }\href {\doibase 10.1103/PhysRevD.26.1296} {\bibfield  {journal} {\bibinfo
  {journal} {Phys. Rev.}\ }\textbf {\bibinfo {volume} {D26}},\ \bibinfo {pages}
  {1296} (\bibinfo {year} {1982})}\BibitemShut {NoStop}%
%\%CITATION = PHRVA,D26,1296;\%\%
\bibitem [{\citenamefont {Adams}\ \emph {et~al.}(1993)\citenamefont {Adams},
  \citenamefont {Fatuzzo}, \citenamefont {Freese}, \citenamefont {Tarle},
  \citenamefont {Watkins},\ and\ \citenamefont {Turner}}]{adams1993extension}%
  \BibitemOpen
  \bibfield  {author} {\bibinfo {author} {\bibfnamefont {F.~C.}\ \bibnamefont
  {Adams}}, \bibinfo {author} {\bibfnamefont {M.}~\bibnamefont {Fatuzzo}},
  \bibinfo {author} {\bibfnamefont {K.}~\bibnamefont {Freese}}, \bibinfo
  {author} {\bibfnamefont {G.}~\bibnamefont {Tarle}}, \bibinfo {author}
  {\bibfnamefont {R.}~\bibnamefont {Watkins}}, \ and\ \bibinfo {author}
  {\bibfnamefont {M.~S.}\ \bibnamefont {Turner}},\ }\href {\doibase
  10.1103/PhysRevLett.70.2511} {\bibfield  {journal} {\bibinfo  {journal}
  {Phys. Rev. Lett.}\ }\textbf {\bibinfo {volume} {70}},\ \bibinfo {pages}
  {2511} (\bibinfo {year} {1993})}\BibitemShut {NoStop}%
%\%CITATION = PRLTA,70,2511;\%\%
\bibitem [{\citenamefont {Ambrosio}\ \emph {et~al.}(2002)\citenamefont
  {Ambrosio} \emph {et~al.}}]{ambrosio2002final}%
  \BibitemOpen
  \bibfield  {author} {\bibinfo {author} {\bibfnamefont {M.}~\bibnamefont
  {Ambrosio}} \emph {et~al.} (\bibinfo {collaboration} {MACRO}),\ }\href
  {\doibase 10.1140/epjc/s2002-01046-9} {\bibfield  {journal} {\bibinfo
  {journal} {Eur. Phys. J.}\ }\textbf {\bibinfo {volume} {C25}},\ \bibinfo
  {pages} {511} (\bibinfo {year} {2002})},\ \Eprint
  {http://arxiv.org/abs/hep-ex/0207020} {arXiv:hep-ex/0207020 [hep-ex]}
  \BibitemShut {NoStop}%
%\%CITATION = HEP-EX/0207020;\%\%
\bibitem [{\citenamefont {Detrixhe}\ \emph {et~al.}(2011)\citenamefont
  {Detrixhe} \emph {et~al.}}]{detrixhe2010ultrarelativistic}%
  \BibitemOpen
  \bibfield  {author} {\bibinfo {author} {\bibfnamefont {M.}~\bibnamefont
  {Detrixhe}} \emph {et~al.} (\bibinfo {collaboration} {ANITA-II}),\ }\href
  {\doibase 10.1103/PhysRevD.83.023513} {\bibfield  {journal} {\bibinfo
  {journal} {Phys. Rev.}\ }\textbf {\bibinfo {volume} {D83}},\ \bibinfo {pages}
  {023513} (\bibinfo {year} {2011})},\ \Eprint {http://arxiv.org/abs/1008.1282}
  {arXiv:1008.1282 [astro-ph.HE]} \BibitemShut {NoStop}%
%\%CITATION = ARXIV:1008.1282;\%\%
\bibitem [{\citenamefont {Adrian-Martinez}\ \emph {et~al.}(2012)\citenamefont
  {Adrian-Martinez} \emph {et~al.}}]{adrianmartinez2011search}%
  \BibitemOpen
  \bibfield  {author} {\bibinfo {author} {\bibfnamefont {S.}~\bibnamefont
  {Adrian-Martinez}} \emph {et~al.} (\bibinfo {collaboration} {ANTARES}),\
  }\href {\doibase 10.1016/j.astropartphys.2012.02.007} {\bibfield  {journal}
  {\bibinfo  {journal} {Astropart. Phys.}\ }\textbf {\bibinfo {volume} {35}},\
  \bibinfo {pages} {634} (\bibinfo {year} {2012})},\ \Eprint
  {http://arxiv.org/abs/1110.2656} {arXiv:1110.2656 [astro-ph.HE]} \BibitemShut
  {NoStop}%
%\%CITATION = ARXIV:1110.2656;\%\%
\bibitem [{\citenamefont {Aartsen}\ \emph {et~al.}(2016)\citenamefont {Aartsen}
  \emph {et~al.}}]{aartsen2015searches}%
  \BibitemOpen
  \bibfield  {author} {\bibinfo {author} {\bibfnamefont {M.~G.}\ \bibnamefont
  {Aartsen}} \emph {et~al.} (\bibinfo {collaboration} {IceCube}),\ }\href
  {\doibase 10.1140/epjc/s10052-016-3953-8} {\bibfield  {journal} {\bibinfo
  {journal} {Eur. Phys. J.}\ }\textbf {\bibinfo {volume} {C76}},\ \bibinfo
  {pages} {133} (\bibinfo {year} {2016})},\ \Eprint
  {http://arxiv.org/abs/1511.01350} {arXiv:1511.01350 [astro-ph.HE]}
  \BibitemShut {NoStop}%
%\%CITATION = ARXIV:1511.01350;\%\%
\bibitem [{\citenamefont {Burdin}\ \emph {et~al.}(2015)\citenamefont {Burdin},
  \citenamefont {Fairbairn}, \citenamefont {Mermod}, \citenamefont {Milstead},
  \citenamefont {Pinfold}, \citenamefont {Sloan},\ and\ \citenamefont
  {Taylor}}]{burdin2014noncollider}%
  \BibitemOpen
  \bibfield  {author} {\bibinfo {author} {\bibfnamefont {S.}~\bibnamefont
  {Burdin}}, \bibinfo {author} {\bibfnamefont {M.}~\bibnamefont {Fairbairn}},
  \bibinfo {author} {\bibfnamefont {P.}~\bibnamefont {Mermod}}, \bibinfo
  {author} {\bibfnamefont {D.}~\bibnamefont {Milstead}}, \bibinfo {author}
  {\bibfnamefont {J.}~\bibnamefont {Pinfold}}, \bibinfo {author} {\bibfnamefont
  {T.}~\bibnamefont {Sloan}}, \ and\ \bibinfo {author} {\bibfnamefont
  {W.}~\bibnamefont {Taylor}},\ }\href {\doibase 10.1016/j.physrep.2015.03.004}
  {\bibfield  {journal} {\bibinfo  {journal} {Phys. Rept.}\ }\textbf {\bibinfo
  {volume} {582}},\ \bibinfo {pages} {1} (\bibinfo {year} {2015})},\ \Eprint
  {http://arxiv.org/abs/1410.1374} {arXiv:1410.1374 [hep-ph]} \BibitemShut
  {NoStop}%
%\%CITATION = ARXIV:1410.1374;\%\%
\bibitem [{\citenamefont {Fairbairn}\ \emph {et~al.}(2007)\citenamefont
  {Fairbairn}, \citenamefont {Kraan}, \citenamefont {Milstead}, \citenamefont
  {Sjostrand}, \citenamefont {Skands},\ and\ \citenamefont
  {Sloan}}]{fairbairn2006stable}%
  \BibitemOpen
  \bibfield  {author} {\bibinfo {author} {\bibfnamefont {M.}~\bibnamefont
  {Fairbairn}}, \bibinfo {author} {\bibfnamefont {A.~C.}\ \bibnamefont
  {Kraan}}, \bibinfo {author} {\bibfnamefont {D.~A.}\ \bibnamefont {Milstead}},
  \bibinfo {author} {\bibfnamefont {T.}~\bibnamefont {Sjostrand}}, \bibinfo
  {author} {\bibfnamefont {P.~Z.}\ \bibnamefont {Skands}}, \ and\ \bibinfo
  {author} {\bibfnamefont {T.}~\bibnamefont {Sloan}},\ }\href {\doibase
  10.1016/j.physrep.2006.10.002} {\bibfield  {journal} {\bibinfo  {journal}
  {Phys. Rept.}\ }\textbf {\bibinfo {volume} {438}},\ \bibinfo {pages} {1}
  (\bibinfo {year} {2007})},\ \Eprint {http://arxiv.org/abs/hep-ph/0611040}
  {arXiv:hep-ph/0611040 [hep-ph]} \BibitemShut {NoStop}%
%\%CITATION = HEP-PH/0611040;\%\%
\bibitem [{\citenamefont {Aad}\ \emph {et~al.}(2016)\citenamefont {Aad} \emph
  {et~al.}}]{aad2015search}%
  \BibitemOpen
  \bibfield  {author} {\bibinfo {author} {\bibfnamefont {G.}~\bibnamefont
  {Aad}} \emph {et~al.} (\bibinfo {collaboration} {ATLAS}),\ }\href {\doibase
  10.1103/PhysRevD.93.052009} {\bibfield  {journal} {\bibinfo  {journal} {Phys.
  Rev.}\ }\textbf {\bibinfo {volume} {D93}},\ \bibinfo {pages} {052009}
  (\bibinfo {year} {2016})},\ \Eprint {http://arxiv.org/abs/1509.08059}
  {arXiv:1509.08059 [hep-ex]} \BibitemShut {NoStop}%
%\%CITATION = ARXIV:1509.08059;\%\%
\bibitem [{\citenamefont {Patrignani}\ \emph {et~al.}(2016)\citenamefont
  {Patrignani} \emph {et~al.}}]{olive2016review}%
  \BibitemOpen
  \bibfield  {author} {\bibinfo {author} {\bibfnamefont {C.}~\bibnamefont
  {Patrignani}} \emph {et~al.} (\bibinfo {collaboration} {Particle Data
  Group}),\ }\href {\doibase 10.1088/1674-1137/40/10/100001} {\bibfield
  {journal} {\bibinfo  {journal} {Chin. Phys.}\ }\textbf {\bibinfo {volume}
  {C40}},\ \bibinfo {pages} {100001} (\bibinfo {year} {2016})}\BibitemShut
  {NoStop}%
%\%CITATION = CHPHD,C40,100001;\%\%
\bibitem [{\citenamefont {Acharya}\ \emph {et~al.}(2017)\citenamefont {Acharya}
  \emph {et~al.}}]{acharya2016search}%
  \BibitemOpen
  \bibfield  {author} {\bibinfo {author} {\bibfnamefont {B.}~\bibnamefont
  {Acharya}} \emph {et~al.} (\bibinfo {collaboration} {MoEDAL}),\ }\href
  {\doibase 10.1103/PhysRevLett.118.061801} {\bibfield  {journal} {\bibinfo
  {journal} {Phys. Rev. Lett.}\ }\textbf {\bibinfo {volume} {118}},\ \bibinfo
  {pages} {061801} (\bibinfo {year} {2017})},\ \Eprint
  {http://arxiv.org/abs/1611.06817} {arXiv:1611.06817 [hep-ex]} \BibitemShut
  {NoStop}%
%\%CITATION = ARXIV:1611.06817;\%\%
\bibitem [{\citenamefont {Englert}\ and\ \citenamefont
  {Jaeckel}(2017)}]{englert2016getting}%
  \BibitemOpen
  \bibfield  {author} {\bibinfo {author} {\bibfnamefont {C.}~\bibnamefont
  {Englert}}\ and\ \bibinfo {author} {\bibfnamefont {J.}~\bibnamefont
  {Jaeckel}},\ }\href {\doibase 10.1016/j.physletb.2017.01.014} {\bibfield
  {journal} {\bibinfo  {journal} {Phys. Lett.}\ }\textbf {\bibinfo {volume}
  {B769}},\ \bibinfo {pages} {513} (\bibinfo {year} {2017})},\ \Eprint
  {http://arxiv.org/abs/1610.06753} {arXiv:1610.06753 [hep-ph]} \BibitemShut
  {NoStop}%
%\%CITATION = ARXIV:1610.06753;\%\%
\bibitem [{\citenamefont {Drukier}\ and\ \citenamefont
  {Nussinov}(1982)}]{drukier1982monopole}%
  \BibitemOpen
  \bibfield  {author} {\bibinfo {author} {\bibfnamefont {A.~K.}\ \bibnamefont
  {Drukier}}\ and\ \bibinfo {author} {\bibfnamefont {S.}~\bibnamefont
  {Nussinov}},\ }\href {\doibase 10.1103/PhysRevLett.49.102} {\bibfield
  {journal} {\bibinfo  {journal} {Phys. Rev. Lett.}\ }\textbf {\bibinfo
  {volume} {49}},\ \bibinfo {pages} {102} (\bibinfo {year} {1982})}\BibitemShut
  {NoStop}%
%\%CITATION = PRLTA,49,102;\%\%
\bibitem [{\citenamefont {Dirac}(1948)}]{dirac1948theory}%
  \BibitemOpen
  \bibfield  {author} {\bibinfo {author} {\bibfnamefont {P.~A.~M.}\
  \bibnamefont {Dirac}},\ }\href {\doibase 10.1103/PhysRev.74.817} {\bibfield
  {journal} {\bibinfo  {journal} {Phys. Rev.}\ }\textbf {\bibinfo {volume}
  {74}},\ \bibinfo {pages} {817} (\bibinfo {year} {1948})}\BibitemShut
  {NoStop}%
%\%CITATION = PHRVA,74,817;\%\%
\bibitem [{\citenamefont {Cabibbo}\ and\ \citenamefont
  {Ferrari}(1962)}]{cabibbo1962quantum}%
  \BibitemOpen
  \bibfield  {author} {\bibinfo {author} {\bibfnamefont {N.}~\bibnamefont
  {Cabibbo}}\ and\ \bibinfo {author} {\bibfnamefont {E.}~\bibnamefont
  {Ferrari}},\ }\href {\doibase 10.1007/BF02731275} {\bibfield  {journal}
  {\bibinfo  {journal} {Nuovo Cim.}\ }\textbf {\bibinfo {volume} {23}},\
  \bibinfo {pages} {1147} (\bibinfo {year} {1962})}\BibitemShut {NoStop}%
%\%CITATION = NUCIA,23,1147;\%\%
\bibitem [{\citenamefont {Schwinger}(1966)}]{schwinger1966magnetic}%
  \BibitemOpen
  \bibfield  {author} {\bibinfo {author} {\bibfnamefont {J.~S.}\ \bibnamefont
  {Schwinger}},\ }\href {\doibase 10.1103/PhysRev.144.1087} {\bibfield
  {journal} {\bibinfo  {journal} {Phys. Rev.}\ }\textbf {\bibinfo {volume}
  {144}},\ \bibinfo {pages} {1087} (\bibinfo {year} {1966})}\BibitemShut
  {NoStop}%
%\%CITATION = PHRVA,144,1087;\%\%
\bibitem [{\citenamefont {Zwanziger}(1971)}]{zwanziger1970local}%
  \BibitemOpen
  \bibfield  {author} {\bibinfo {author} {\bibfnamefont {D.}~\bibnamefont
  {Zwanziger}},\ }\href {\doibase 10.1103/PhysRevD.3.880} {\bibfield  {journal}
  {\bibinfo  {journal} {Phys. Rev.}\ }\textbf {\bibinfo {volume} {D3}},\
  \bibinfo {pages} {880} (\bibinfo {year} {1971})}\BibitemShut {NoStop}%
%\%CITATION = PHRVA,D3,880;\%\%
\bibitem [{\citenamefont {Blagojevic}\ and\ \citenamefont
  {Senjanovic}(1988)}]{blagojevic1985quantum}%
  \BibitemOpen
  \bibfield  {author} {\bibinfo {author} {\bibfnamefont {M.}~\bibnamefont
  {Blagojevic}}\ and\ \bibinfo {author} {\bibfnamefont {P.}~\bibnamefont
  {Senjanovic}},\ }\href {\doibase 10.1016/0370-1573(88)90098-1} {\bibfield
  {journal} {\bibinfo  {journal} {Phys. Rept.}\ }\textbf {\bibinfo {volume}
  {157}},\ \bibinfo {pages} {233} (\bibinfo {year} {1988})}\BibitemShut
  {NoStop}%
%\%CITATION = PRPLC,157,233;\%\%
\bibitem [{\citenamefont {Milton}(2006)}]{milton2006theoretical}%
  \BibitemOpen
  \bibfield  {author} {\bibinfo {author} {\bibfnamefont {K.~A.}\ \bibnamefont
  {Milton}},\ }\href {\doibase 10.1088/0034-4885/69/6/R02} {\bibfield
  {journal} {\bibinfo  {journal} {Rept. Prog. Phys.}\ }\textbf {\bibinfo
  {volume} {69}},\ \bibinfo {pages} {1637} (\bibinfo {year} {2006})},\ \Eprint
  {http://arxiv.org/abs/hep-ex/0602040} {arXiv:hep-ex/0602040 [hep-ex]}
  \BibitemShut {NoStop}%
%\%CITATION = HEP-EX/0602040;\%\%
\bibitem [{\citenamefont {Gould}\ and\ \citenamefont
  {Rajantie}(2017)}]{gould2017thermal}%
  \BibitemOpen
  \bibfield  {author} {\bibinfo {author} {\bibfnamefont {O.}~\bibnamefont
  {Gould}}\ and\ \bibinfo {author} {\bibfnamefont {A.}~\bibnamefont
  {Rajantie}},\ }\href@noop {} {\  (\bibinfo {year} {2017})},\ \Eprint
  {http://arxiv.org/abs/1704.04801} {arXiv:1704.04801 [hep-th]} \BibitemShut
  {NoStop}%
%\%CITATION = ARXIV:1704.04801;\%\%
\bibitem [{\citenamefont {Affleck}\ and\ \citenamefont
  {Manton}(1982)}]{affleck1981monopole}%
  \BibitemOpen
  \bibfield  {author} {\bibinfo {author} {\bibfnamefont {I.~K.}\ \bibnamefont
  {Affleck}}\ and\ \bibinfo {author} {\bibfnamefont {N.~S.}\ \bibnamefont
  {Manton}},\ }\href@noop {} {\bibfield  {journal} {\bibinfo  {journal}
  {Nuclear Physics B}\ }\textbf {\bibinfo {volume} {194}},\ \bibinfo {pages}
  {38} (\bibinfo {year} {1982})}\BibitemShut {NoStop}%
\bibitem [{\citenamefont {Affleck}\ \emph {et~al.}(1982)\citenamefont
  {Affleck}, \citenamefont {Alvarez},\ and\ \citenamefont
  {Manton}}]{affleck1981pair}%
  \BibitemOpen
  \bibfield  {author} {\bibinfo {author} {\bibfnamefont {I.~K.}\ \bibnamefont
  {Affleck}}, \bibinfo {author} {\bibfnamefont {O.}~\bibnamefont {Alvarez}}, \
  and\ \bibinfo {author} {\bibfnamefont {N.~S.}\ \bibnamefont {Manton}},\
  }\href {\doibase 10.1016/0550-3213(82)90455-2} {\bibfield  {journal}
  {\bibinfo  {journal} {Nucl. Phys.}\ }\textbf {\bibinfo {volume} {B197}},\
  \bibinfo {pages} {509} (\bibinfo {year} {1982})}\BibitemShut {NoStop}%
%\%CITATION = NUPHA,B197,509;\%\%
\bibitem [{\citenamefont {{'t
  Hooft}}(1976{\natexlab{a}})}]{thooft1976symmetry}%
  \BibitemOpen
  \bibfield  {author} {\bibinfo {author} {\bibfnamefont {G.}~\bibnamefont {{'t
  Hooft}}},\ }\href {\doibase 10.1103/PhysRevLett.37.8} {\bibfield  {journal}
  {\bibinfo  {journal} {Phys. Rev. Lett.}\ }\textbf {\bibinfo {volume} {37}},\
  \bibinfo {pages} {8} (\bibinfo {year} {1976}{\natexlab{a}})}\BibitemShut
  {NoStop}%
%\%CITATION = PRLTA,37,8;\%\%
\bibitem [{\citenamefont {{'t
  Hooft}}(1976{\natexlab{b}})}]{thooft1976computation}%
  \BibitemOpen
  \bibfield  {author} {\bibinfo {author} {\bibfnamefont {G.}~\bibnamefont {{'t
  Hooft}}},\ }\href {\doibase
  10.1103/PhysRevD.18.2199.3;;;;;;;;;;;;;;;;;;;;;;;;;;;;;;;;;;;;;;;;;;;;;;;;;;
  10.1103/PhysRevD.14.3432} {\bibfield  {journal} {\bibinfo  {journal} {Phys.
  Rev.}\ }\textbf {\bibinfo {volume} {D14}},\ \bibinfo {pages} {3432} (\bibinfo
  {year} {1976}{\natexlab{b}})},\ \bibinfo {note} {[Erratum: Phys.
  Rev.D18,2199(1978)]}\BibitemShut {NoStop}%
%\%CITATION = PHRVA,D14,3432;\%\%
\bibitem [{\citenamefont {Klinkhamer}\ and\ \citenamefont
  {Manton}(1984)}]{klinkhamer1984saddle}%
  \BibitemOpen
  \bibfield  {author} {\bibinfo {author} {\bibfnamefont {F.~R.}\ \bibnamefont
  {Klinkhamer}}\ and\ \bibinfo {author} {\bibfnamefont {N.~S.}\ \bibnamefont
  {Manton}},\ }\href {\doibase 10.1103/PhysRevD.30.2212} {\bibfield  {journal}
  {\bibinfo  {journal} {Phys. Rev.}\ }\textbf {\bibinfo {volume} {D30}},\
  \bibinfo {pages} {2212} (\bibinfo {year} {1984})}\BibitemShut {NoStop}%
%\%CITATION = PHRVA,D30,2212;\%\%
\bibitem [{\citenamefont {Kuzmin}\ \emph {et~al.}(1985)\citenamefont {Kuzmin},
  \citenamefont {Rubakov},\ and\ \citenamefont
  {Shaposhnikov}}]{kuzmin1985anomalous}%
  \BibitemOpen
  \bibfield  {author} {\bibinfo {author} {\bibfnamefont {V.~A.}\ \bibnamefont
  {Kuzmin}}, \bibinfo {author} {\bibfnamefont {V.~A.}\ \bibnamefont {Rubakov}},
  \ and\ \bibinfo {author} {\bibfnamefont {M.~E.}\ \bibnamefont
  {Shaposhnikov}},\ }\href {\doibase 10.1016/0370-2693(85)91028-7} {\bibfield
  {journal} {\bibinfo  {journal} {Phys. Lett.}\ }\textbf {\bibinfo {volume}
  {155B}},\ \bibinfo {pages} {36} (\bibinfo {year} {1985})}\BibitemShut
  {NoStop}%
%\%CITATION = PHLTA,155B,36;\%\%
\bibitem [{\citenamefont {Arnold}\ and\ \citenamefont
  {McLerran}(1987)}]{arnold1987sphalerons}%
  \BibitemOpen
  \bibfield  {author} {\bibinfo {author} {\bibfnamefont {P.~B.}\ \bibnamefont
  {Arnold}}\ and\ \bibinfo {author} {\bibfnamefont {L.~D.}\ \bibnamefont
  {McLerran}},\ }\href {\doibase 10.1103/PhysRevD.36.581} {\bibfield  {journal}
  {\bibinfo  {journal} {Phys. Rev.}\ }\textbf {\bibinfo {volume} {D36}},\
  \bibinfo {pages} {581} (\bibinfo {year} {1987})}\BibitemShut {NoStop}%
%\%CITATION = PHRVA,D36,581;\%\%
\bibitem [{\citenamefont {Arnold}\ and\ \citenamefont
  {McLerran}(1988)}]{arnold1987sphaleron}%
  \BibitemOpen
  \bibfield  {author} {\bibinfo {author} {\bibfnamefont {P.~B.}\ \bibnamefont
  {Arnold}}\ and\ \bibinfo {author} {\bibfnamefont {L.~D.}\ \bibnamefont
  {McLerran}},\ }\href {\doibase 10.1103/PhysRevD.37.1020} {\bibfield
  {journal} {\bibinfo  {journal} {Phys. Rev.}\ }\textbf {\bibinfo {volume}
  {D37}},\ \bibinfo {pages} {1020} (\bibinfo {year} {1988})}\BibitemShut
  {NoStop}%
%\%CITATION = PHRVA,D37,1020;\%\%
\bibitem [{\citenamefont {He}(1997)}]{he1997search}%
  \BibitemOpen
  \bibfield  {author} {\bibinfo {author} {\bibfnamefont {Y.~D.}\ \bibnamefont
  {He}},\ }\href {\doibase 10.1103/PhysRevLett.79.3134} {\bibfield  {journal}
  {\bibinfo  {journal} {Phys. Rev. Lett.}\ }\textbf {\bibinfo {volume} {79}},\
  \bibinfo {pages} {3134} (\bibinfo {year} {1997})}\BibitemShut {NoStop}%
%\%CITATION = PRLTA,79,3134;\%\%
\bibitem [{\citenamefont {Zeldovich}\ and\ \citenamefont
  {Khlopov}(1978)}]{zeldovich1978concentration}%
  \BibitemOpen
  \bibfield  {author} {\bibinfo {author} {\bibfnamefont {{\relax Ya}.~B.}\
  \bibnamefont {Zeldovich}}\ and\ \bibinfo {author} {\bibfnamefont {M.~{\relax
  Yu}.}\ \bibnamefont {Khlopov}},\ }\href {\doibase
  10.1016/0370-2693(78)90232-0} {\bibfield  {journal} {\bibinfo  {journal}
  {Phys. Lett.}\ }\textbf {\bibinfo {volume} {B79}},\ \bibinfo {pages} {239}
  (\bibinfo {year} {1978})}\BibitemShut {NoStop}%
%\%CITATION = PHLTA,B79,239;\%\%
\bibitem [{\citenamefont {Guth}\ and\ \citenamefont
  {Tye}(1980)}]{guth1979phase}%
  \BibitemOpen
  \bibfield  {author} {\bibinfo {author} {\bibfnamefont {A.~H.}\ \bibnamefont
  {Guth}}\ and\ \bibinfo {author} {\bibfnamefont {S.~H.~H.}\ \bibnamefont
  {Tye}},\ }\href {\doibase 10.1103/PhysRevLett.44.631} {\bibfield  {journal}
  {\bibinfo  {journal} {Phys. Rev. Lett.}\ }\textbf {\bibinfo {volume} {44}},\
  \bibinfo {pages} {631} (\bibinfo {year} {1980})},\ \bibinfo {note} {[Erratum:
  Phys. Rev. Lett.44,963(1980)]}\BibitemShut {NoStop}%
%\%CITATION = PRLTA,44,631;\%\%
\bibitem [{\citenamefont {Preskill}(1979)}]{preskill1979cosmological}%
  \BibitemOpen
  \bibfield  {author} {\bibinfo {author} {\bibfnamefont {J.}~\bibnamefont
  {Preskill}},\ }\href {\doibase 10.1103/PhysRevLett.43.1365} {\bibfield
  {journal} {\bibinfo  {journal} {Phys. Rev. Lett.}\ }\textbf {\bibinfo
  {volume} {43}},\ \bibinfo {pages} {1365} (\bibinfo {year}
  {1979})}\BibitemShut {NoStop}%
%\%CITATION = PRLTA,43,1365;\%\%
\bibitem [{\citenamefont {Turner}(1982)}]{turner1982thermal}%
  \BibitemOpen
  \bibfield  {author} {\bibinfo {author} {\bibfnamefont {M.~S.}\ \bibnamefont
  {Turner}},\ }\href {\doibase 10.1016/0370-2693(82)90803-6} {\bibfield
  {journal} {\bibinfo  {journal} {Phys. Lett.}\ }\textbf {\bibinfo {volume}
  {B115}},\ \bibinfo {pages} {95} (\bibinfo {year} {1982})}\BibitemShut
  {NoStop}%
%\%CITATION = PHLTA,B115,95;\%\%
\bibitem [{\citenamefont {Collins}\ and\ \citenamefont
  {Turner}(1984)}]{collins1984thermal}%
  \BibitemOpen
  \bibfield  {author} {\bibinfo {author} {\bibfnamefont {W.}~\bibnamefont
  {Collins}}\ and\ \bibinfo {author} {\bibfnamefont {M.~S.}\ \bibnamefont
  {Turner}},\ }\href {\doibase 10.1103/PhysRevD.29.2158} {\bibfield  {journal}
  {\bibinfo  {journal} {Phys. Rev.}\ }\textbf {\bibinfo {volume} {D29}},\
  \bibinfo {pages} {2158} (\bibinfo {year} {1984})}\BibitemShut {NoStop}%
%\%CITATION = PHRVA,D29,2158;\%\%
\bibitem [{\citenamefont {Lindblom}\ and\ \citenamefont
  {Steinhardt}(1985)}]{lindblom1984thermal}%
  \BibitemOpen
  \bibfield  {author} {\bibinfo {author} {\bibfnamefont {P.~R.}\ \bibnamefont
  {Lindblom}}\ and\ \bibinfo {author} {\bibfnamefont {P.~J.}\ \bibnamefont
  {Steinhardt}},\ }\href {\doibase 10.1103/PhysRevD.31.2151} {\bibfield
  {journal} {\bibinfo  {journal} {Phys. Rev.}\ }\textbf {\bibinfo {volume}
  {D31}},\ \bibinfo {pages} {2151} (\bibinfo {year} {1985})}\BibitemShut
  {NoStop}%
%\%CITATION = PHRVA,D31,2151;\%\%
\bibitem [{\citenamefont {Boulware}\ \emph {et~al.}(1976)\citenamefont
  {Boulware}, \citenamefont {Brown}, \citenamefont {Cahn}, \citenamefont
  {Ellis},\ and\ \citenamefont {Lee}}]{boulware1976scattering}%
  \BibitemOpen
  \bibfield  {author} {\bibinfo {author} {\bibfnamefont {D.~G.}\ \bibnamefont
  {Boulware}}, \bibinfo {author} {\bibfnamefont {L.~S.}\ \bibnamefont {Brown}},
  \bibinfo {author} {\bibfnamefont {R.~N.}\ \bibnamefont {Cahn}}, \bibinfo
  {author} {\bibfnamefont {S.~D.}\ \bibnamefont {Ellis}}, \ and\ \bibinfo
  {author} {\bibfnamefont {C.-k.}\ \bibnamefont {Lee}},\ }\href {\doibase
  10.1103/PhysRevD.14.2708} {\bibfield  {journal} {\bibinfo  {journal} {Phys.
  Rev.}\ }\textbf {\bibinfo {volume} {D14}},\ \bibinfo {pages} {2708} (\bibinfo
  {year} {1976})}\BibitemShut {NoStop}%
%\%CITATION = PHRVA,D14,2708;\%\%
\bibitem [{\citenamefont {Bardakci}\ and\ \citenamefont
  {Samuel}(1978)}]{bardakci1978local}%
  \BibitemOpen
  \bibfield  {author} {\bibinfo {author} {\bibfnamefont {K.}~\bibnamefont
  {Bardakci}}\ and\ \bibinfo {author} {\bibfnamefont {S.}~\bibnamefont
  {Samuel}},\ }\href {\doibase 10.1103/PhysRevD.18.2849} {\bibfield  {journal}
  {\bibinfo  {journal} {Phys. Rev.}\ }\textbf {\bibinfo {volume} {D18}},\
  \bibinfo {pages} {2849} (\bibinfo {year} {1978})}\BibitemShut {NoStop}%
%\%CITATION = PHRVA,D18,2849;\%\%
\bibitem [{\citenamefont {Manton}(1979)}]{manton1978effective}%
  \BibitemOpen
  \bibfield  {author} {\bibinfo {author} {\bibfnamefont {N.~S.}\ \bibnamefont
  {Manton}},\ }\href {\doibase 10.1016/0550-3213(79)90309-2} {\bibfield
  {journal} {\bibinfo  {journal} {Nucl. Phys.}\ }\textbf {\bibinfo {volume}
  {B150}},\ \bibinfo {pages} {397} (\bibinfo {year} {1979})}\BibitemShut
  {NoStop}%
%\%CITATION = NUPHA,B150,397;\%\%
\bibitem [{\citenamefont {Callan}\ and\ \citenamefont
  {Coleman}(1977)}]{callan1977fate}%
  \BibitemOpen
  \bibfield  {author} {\bibinfo {author} {\bibfnamefont {J.~C.~G.}\
  \bibnamefont {Callan}}\ and\ \bibinfo {author} {\bibfnamefont {S.~R.}\
  \bibnamefont {Coleman}},\ }\href {\doibase 10.1103/PhysRevD.16.1762}
  {\bibfield  {journal} {\bibinfo  {journal} {Phys. Rev.}\ }\textbf {\bibinfo
  {volume} {D16}},\ \bibinfo {pages} {1762} (\bibinfo {year}
  {1977})}\BibitemShut {NoStop}%
%\%CITATION = PHRVA,D16,1762;\%\%
\bibitem [{\citenamefont {Kharzeev}\ \emph {et~al.}(2008)\citenamefont
  {Kharzeev}, \citenamefont {McLerran},\ and\ \citenamefont
  {Warringa}}]{kharzeev2007effects}%
  \BibitemOpen
  \bibfield  {author} {\bibinfo {author} {\bibfnamefont {D.~E.}\ \bibnamefont
  {Kharzeev}}, \bibinfo {author} {\bibfnamefont {L.~D.}\ \bibnamefont
  {McLerran}}, \ and\ \bibinfo {author} {\bibfnamefont {H.~J.}\ \bibnamefont
  {Warringa}},\ }\href {\doibase 10.1016/j.nuclphysa.2008.02.298} {\bibfield
  {journal} {\bibinfo  {journal} {Nucl. Phys.}\ }\textbf {\bibinfo {volume}
  {A803}},\ \bibinfo {pages} {227} (\bibinfo {year} {2008})},\ \Eprint
  {http://arxiv.org/abs/0711.0950} {arXiv:0711.0950 [hep-ph]} \BibitemShut
  {NoStop}%
%\%CITATION = ARXIV:0711.0950;\%\%
\bibitem [{\citenamefont {Skokov}\ \emph {et~al.}(2009)\citenamefont {Skokov},
  \citenamefont {Illarionov},\ and\ \citenamefont
  {Toneev}}]{skokov2009estimate}%
  \BibitemOpen
  \bibfield  {author} {\bibinfo {author} {\bibfnamefont {V.}~\bibnamefont
  {Skokov}}, \bibinfo {author} {\bibfnamefont {A.~{\relax Yu}.}\ \bibnamefont
  {Illarionov}}, \ and\ \bibinfo {author} {\bibfnamefont {V.}~\bibnamefont
  {Toneev}},\ }\href {\doibase 10.1142/S0217751X09047570} {\bibfield  {journal}
  {\bibinfo  {journal} {Int. J. Mod. Phys.}\ }\textbf {\bibinfo {volume}
  {A24}},\ \bibinfo {pages} {5925} (\bibinfo {year} {2009})},\ \Eprint
  {http://arxiv.org/abs/0907.1396} {arXiv:0907.1396 [nucl-th]} \BibitemShut
  {NoStop}%
%\%CITATION = ARXIV:0907.1396;\%\%
\bibitem [{\citenamefont {Voronyuk}\ \emph {et~al.}(2011)\citenamefont
  {Voronyuk}, \citenamefont {Toneev}, \citenamefont {Cassing}, \citenamefont
  {Bratkovskaya}, \citenamefont {Konchakovski},\ and\ \citenamefont
  {Voloshin}}]{voronyuk2011electromagnetic}%
  \BibitemOpen
  \bibfield  {author} {\bibinfo {author} {\bibfnamefont {V.}~\bibnamefont
  {Voronyuk}}, \bibinfo {author} {\bibfnamefont {V.~D.}\ \bibnamefont
  {Toneev}}, \bibinfo {author} {\bibfnamefont {W.}~\bibnamefont {Cassing}},
  \bibinfo {author} {\bibfnamefont {E.~L.}\ \bibnamefont {Bratkovskaya}},
  \bibinfo {author} {\bibfnamefont {V.~P.}\ \bibnamefont {Konchakovski}}, \
  and\ \bibinfo {author} {\bibfnamefont {S.~A.}\ \bibnamefont {Voloshin}},\
  }\href {\doibase 10.1103/PhysRevC.83.054911} {\bibfield  {journal} {\bibinfo
  {journal} {Phys. Rev.}\ }\textbf {\bibinfo {volume} {C83}},\ \bibinfo {pages}
  {054911} (\bibinfo {year} {2011})},\ \Eprint {http://arxiv.org/abs/1103.4239}
  {arXiv:1103.4239 [nucl-th]} \BibitemShut {NoStop}%
%\%CITATION = ARXIV:1103.4239;\%\%
\bibitem [{\citenamefont {Bzdak}\ and\ \citenamefont
  {Skokov}(2012)}]{bzdak2011event}%
  \BibitemOpen
  \bibfield  {author} {\bibinfo {author} {\bibfnamefont {A.}~\bibnamefont
  {Bzdak}}\ and\ \bibinfo {author} {\bibfnamefont {V.}~\bibnamefont {Skokov}},\
  }\href {\doibase 10.1016/j.physletb.2012.02.065} {\bibfield  {journal}
  {\bibinfo  {journal} {Phys. Lett.}\ }\textbf {\bibinfo {volume} {B710}},\
  \bibinfo {pages} {171} (\bibinfo {year} {2012})},\ \Eprint
  {http://arxiv.org/abs/1111.1949} {arXiv:1111.1949 [hep-ph]} \BibitemShut
  {NoStop}%
%\%CITATION = ARXIV:1111.1949;\%\%
\bibitem [{\citenamefont {Deng}\ and\ \citenamefont
  {Huang}(2012)}]{deng2012event}%
  \BibitemOpen
  \bibfield  {author} {\bibinfo {author} {\bibfnamefont {W.-T.}\ \bibnamefont
  {Deng}}\ and\ \bibinfo {author} {\bibfnamefont {X.-G.}\ \bibnamefont
  {Huang}},\ }\href {\doibase 10.1103/PhysRevC.85.044907} {\bibfield  {journal}
  {\bibinfo  {journal} {Phys. Rev.}\ }\textbf {\bibinfo {volume} {C85}},\
  \bibinfo {pages} {044907} (\bibinfo {year} {2012})},\ \Eprint
  {http://arxiv.org/abs/1201.5108} {arXiv:1201.5108 [nucl-th]} \BibitemShut
  {NoStop}%
%\%CITATION = ARXIV:1201.5108;\%\%
\bibitem [{\citenamefont {McLerran}\ and\ \citenamefont
  {Skokov}(2014)}]{mclerran2013comments}%
  \BibitemOpen
  \bibfield  {author} {\bibinfo {author} {\bibfnamefont {L.}~\bibnamefont
  {McLerran}}\ and\ \bibinfo {author} {\bibfnamefont {V.}~\bibnamefont
  {Skokov}},\ }\href {\doibase 10.1016/j.nuclphysa.2014.05.008} {\bibfield
  {journal} {\bibinfo  {journal} {Nucl. Phys.}\ }\textbf {\bibinfo {volume}
  {A929}},\ \bibinfo {pages} {184} (\bibinfo {year} {2014})},\ \Eprint
  {http://arxiv.org/abs/1305.0774} {arXiv:1305.0774 [hep-ph]} \BibitemShut
  {NoStop}%
%\%CITATION = ARXIV:1305.0774;\%\%
\bibitem [{\citenamefont {{De Roeck}}\ \emph {et~al.}(2012)\citenamefont {{De
  Roeck}}, \citenamefont {H{\"a}chler}, \citenamefont {Hirt}, \citenamefont
  {Joergensen}, \citenamefont {Katre}, \citenamefont {Mermod}, \citenamefont
  {Milstead},\ and\ \citenamefont {Sloan}}]{deroeck2012development}%
  \BibitemOpen
  \bibfield  {author} {\bibinfo {author} {\bibfnamefont {A.}~\bibnamefont {{De
  Roeck}}}, \bibinfo {author} {\bibfnamefont {H.~P.}\ \bibnamefont
  {H{\"a}chler}}, \bibinfo {author} {\bibfnamefont {A.~M.}\ \bibnamefont
  {Hirt}}, \bibinfo {author} {\bibfnamefont {M.~D.}\ \bibnamefont
  {Joergensen}}, \bibinfo {author} {\bibfnamefont {A.}~\bibnamefont {Katre}},
  \bibinfo {author} {\bibfnamefont {P.}~\bibnamefont {Mermod}}, \bibinfo
  {author} {\bibfnamefont {D.}~\bibnamefont {Milstead}}, \ and\ \bibinfo
  {author} {\bibfnamefont {T.}~\bibnamefont {Sloan}},\ }\href {\doibase
  10.1140/epjc} {\bibfield  {journal} {\bibinfo  {journal} {Eur. Phys. J.}\
  }\textbf {\bibinfo {volume} {C72}},\ \bibinfo {pages} {2212} (\bibinfo {year}
  {2012})}\BibitemShut {NoStop}%
%\%CITATION = EPHJA,C72,2212;\%\%
\bibitem [{\citenamefont {Joergensen}\ \emph {et~al.}(2012)\citenamefont
  {Joergensen}, \citenamefont {{De Roeck}}, \citenamefont {Hachler},
  \citenamefont {Hirt}, \citenamefont {Katre}, \citenamefont {Mermod},
  \citenamefont {Milstead},\ and\ \citenamefont
  {Sloan}}]{joergensen2012searching}%
  \BibitemOpen
  \bibfield  {author} {\bibinfo {author} {\bibfnamefont {M.~D.}\ \bibnamefont
  {Joergensen}}, \bibinfo {author} {\bibfnamefont {A.}~\bibnamefont {{De
  Roeck}}}, \bibinfo {author} {\bibfnamefont {H.~P.}\ \bibnamefont {Hachler}},
  \bibinfo {author} {\bibfnamefont {A.}~\bibnamefont {Hirt}}, \bibinfo {author}
  {\bibfnamefont {A.}~\bibnamefont {Katre}}, \bibinfo {author} {\bibfnamefont
  {P.}~\bibnamefont {Mermod}}, \bibinfo {author} {\bibfnamefont
  {D.}~\bibnamefont {Milstead}}, \ and\ \bibinfo {author} {\bibfnamefont
  {T.}~\bibnamefont {Sloan}},\ }\href@noop {} {\  (\bibinfo {year} {2012})},\
  \Eprint {http://arxiv.org/abs/1206.6793} {arXiv:1206.6793 [physics.ins-det]}
  \BibitemShut {NoStop}%
%\%CITATION = ARXIV:1206.6793;\%\%
\bibitem [{\citenamefont {Schlagheck}(2000)}]{schlagheck1999thermalization}%
  \BibitemOpen
  \bibfield  {author} {\bibinfo {author} {\bibfnamefont {H.}~\bibnamefont
  {Schlagheck}} (\bibinfo {collaboration} {WA98}),\ }\bibfield  {booktitle}
  {\emph {\bibinfo {booktitle} {{Particles and nuclei. Proceedings, 15th
  International Conference, PANIC '99, Uppsala, Sweden, June 10-16, 1999}}},\
  }\href {\doibase 10.1016/S0375-9474(99)00703-4} {\bibfield  {journal}
  {\bibinfo  {journal} {Nucl. Phys.}\ }\textbf {\bibinfo {volume} {A663}},\
  \bibinfo {pages} {725} (\bibinfo {year} {2000})},\ \Eprint
  {http://arxiv.org/abs/nucl-ex/9909005} {arXiv:nucl-ex/9909005 [nucl-ex]}
  \BibitemShut {NoStop}%
%\%CITATION = NUCL-EX/9909005;\%\%
\bibitem [{\citenamefont {Aggarwal}\ \emph {et~al.}(2000)\citenamefont
  {Aggarwal} \emph {et~al.}}]{aggarwal2000observation}%
  \BibitemOpen
  \bibfield  {author} {\bibinfo {author} {\bibfnamefont {M.~M.}\ \bibnamefont
  {Aggarwal}} \emph {et~al.} (\bibinfo {collaboration} {WA98}),\ }\href
  {\doibase 10.1103/PhysRevLett.85.3595} {\bibfield  {journal} {\bibinfo
  {journal} {Phys. Rev. Lett.}\ }\textbf {\bibinfo {volume} {85}},\ \bibinfo
  {pages} {3595} (\bibinfo {year} {2000})},\ \Eprint
  {http://arxiv.org/abs/nucl-ex/0006008} {arXiv:nucl-ex/0006008 [nucl-ex]}
  \BibitemShut {NoStop}%
%\%CITATION = NUCL-EX/0006008;\%\%
\bibitem [{\citenamefont {Tuchin}(2010)}]{tuchin2010synchrotron}%
  \BibitemOpen
  \bibfield  {author} {\bibinfo {author} {\bibfnamefont {K.}~\bibnamefont
  {Tuchin}},\ }\href {\doibase 10.1103/PhysRevC.83.039903;;;;;;;;;;;;;;;;;;
  10.1103/PhysRevC.82.034904} {\bibfield  {journal} {\bibinfo  {journal} {Phys.
  Rev.}\ }\textbf {\bibinfo {volume} {C82}},\ \bibinfo {pages} {034904}
  (\bibinfo {year} {2010})},\ \bibinfo {note} {[Erratum: Phys.
  Rev.C83,039903(2011)]},\ \Eprint {http://arxiv.org/abs/1006.3051}
  {arXiv:1006.3051 [nucl-th]} \BibitemShut {NoStop}%
%\%CITATION = ARXIV:1006.3051;\%\%
\bibitem [{\citenamefont {Tuchin}(2013{\natexlab{a}})}]{tuchin2013particle}%
  \BibitemOpen
  \bibfield  {author} {\bibinfo {author} {\bibfnamefont {K.}~\bibnamefont
  {Tuchin}},\ }\href {\doibase 10.1155/2013/490495} {\bibfield  {journal}
  {\bibinfo  {journal} {Adv. High Energy Phys.}\ }\textbf {\bibinfo {volume}
  {2013}},\ \bibinfo {pages} {490495} (\bibinfo {year} {2013}{\natexlab{a}})},\
  \Eprint {http://arxiv.org/abs/1301.0099} {arXiv:1301.0099 [hep-ph]}
  \BibitemShut {NoStop}%
%\%CITATION = ARXIV:1301.0099;\%\%
\bibitem [{\citenamefont {Tuchin}(2013{\natexlab{b}})}]{tuchin2013time}%
  \BibitemOpen
  \bibfield  {author} {\bibinfo {author} {\bibfnamefont {K.}~\bibnamefont
  {Tuchin}},\ }\href {\doibase 10.1103/PhysRevC.88.024911} {\bibfield
  {journal} {\bibinfo  {journal} {Phys. Rev.}\ }\textbf {\bibinfo {volume}
  {C88}},\ \bibinfo {pages} {024911} (\bibinfo {year} {2013}{\natexlab{b}})},\
  \Eprint {http://arxiv.org/abs/1305.5806} {arXiv:1305.5806 [hep-ph]}
  \BibitemShut {NoStop}%
%\%CITATION = ARXIV:1305.5806;\%\%
\bibitem [{\citenamefont {Hwa}\ and\ \citenamefont
  {Kajantie}(1986)}]{hwa1985initial}%
  \BibitemOpen
  \bibfield  {author} {\bibinfo {author} {\bibfnamefont {R.~C.}\ \bibnamefont
  {Hwa}}\ and\ \bibinfo {author} {\bibfnamefont {K.}~\bibnamefont {Kajantie}},\
  }\href {\doibase 10.1103/PhysRevLett.56.696} {\bibfield  {journal} {\bibinfo
  {journal} {Phys. Rev. Lett.}\ }\textbf {\bibinfo {volume} {56}},\ \bibinfo
  {pages} {696} (\bibinfo {year} {1986})}\BibitemShut {NoStop}%
%\%CITATION = PRLTA,56,696;\%\%
\bibitem [{\citenamefont {Brezin}\ and\ \citenamefont
  {Itzykson}(1970)}]{brezin1970pair}%
  \BibitemOpen
  \bibfield  {author} {\bibinfo {author} {\bibfnamefont {E.}~\bibnamefont
  {Brezin}}\ and\ \bibinfo {author} {\bibfnamefont {C.}~\bibnamefont
  {Itzykson}},\ }\href {\doibase 10.1103/PhysRevD.2.1191} {\bibfield  {journal}
  {\bibinfo  {journal} {Phys. Rev.}\ }\textbf {\bibinfo {volume} {D2}},\
  \bibinfo {pages} {1191} (\bibinfo {year} {1970})}\BibitemShut {NoStop}%
%\%CITATION = PHRVA,D2,1191;\%\%
\bibitem [{\citenamefont {Dunne}\ and\ \citenamefont
  {Schubert}(2005)}]{dunne2005worldline}%
  \BibitemOpen
  \bibfield  {author} {\bibinfo {author} {\bibfnamefont {G.~V.}\ \bibnamefont
  {Dunne}}\ and\ \bibinfo {author} {\bibfnamefont {C.}~\bibnamefont
  {Schubert}},\ }\href {\doibase 10.1103/PhysRevD.72.105004} {\bibfield
  {journal} {\bibinfo  {journal} {Phys. Rev.}\ }\textbf {\bibinfo {volume}
  {D72}},\ \bibinfo {pages} {105004} (\bibinfo {year} {2005})},\ \Eprint
  {http://arxiv.org/abs/hep-th/0507174} {arXiv:hep-th/0507174 [hep-th]}
  \BibitemShut {NoStop}%
%\%CITATION = HEP-TH/0507174;\%\%
\bibitem [{\citenamefont {Ilderton}\ \emph {et~al.}(2015)\citenamefont
  {Ilderton}, \citenamefont {Torgrimsson},\ and\ \citenamefont
  {W{\aa}rdh}}]{ilderton2015nonperturbative}%
  \BibitemOpen
  \bibfield  {author} {\bibinfo {author} {\bibfnamefont {A.}~\bibnamefont
  {Ilderton}}, \bibinfo {author} {\bibfnamefont {G.}~\bibnamefont
  {Torgrimsson}}, \ and\ \bibinfo {author} {\bibfnamefont {J.}~\bibnamefont
  {W{\aa}rdh}},\ }\href {\doibase 10.1103/PhysRevD.92.065001} {\bibfield
  {journal} {\bibinfo  {journal} {Phys. Rev.}\ }\textbf {\bibinfo {volume}
  {D92}},\ \bibinfo {pages} {065001} (\bibinfo {year} {2015})},\ \Eprint
  {http://arxiv.org/abs/1506.09186} {arXiv:1506.09186 [hep-th]} \BibitemShut
  {NoStop}%
%\%CITATION = ARXIV:1506.09186;\%\%
\bibitem [{\citenamefont {Kohlf{\"u}rst}(2015)}]{kohlfurst2015electron}%
  \BibitemOpen
  \bibfield  {author} {\bibinfo {author} {\bibfnamefont {C.}~\bibnamefont
  {Kohlf{\"u}rst}},\ }\emph {\bibinfo {title} {{Electron-positron pair
  production in inhomogeneous electromagnetic fields}}},\ \href
  {http://inspirehep.net/record/1410605/files/arXiv:1512.06082.pdf} {Ph.D.
  thesis},\ \bibinfo  {school} {U. Graz (main)} (\bibinfo {year} {2015}),\
  \Eprint {http://arxiv.org/abs/1512.06082} {arXiv:1512.06082 [hep-ph]}
  \BibitemShut {NoStop}%
%\%CITATION = ARXIV:1512.06082;\%\%
\bibitem [{\citenamefont {Torgrimsson}\ \emph {et~al.}(2017)\citenamefont
  {Torgrimsson}, \citenamefont {Schneider}, \citenamefont {Oertel},\ and\
  \citenamefont {Sch{\"u}tzhold}}]{torgrimsson2017dynamically}%
  \BibitemOpen
  \bibfield  {author} {\bibinfo {author} {\bibfnamefont {G.}~\bibnamefont
  {Torgrimsson}}, \bibinfo {author} {\bibfnamefont {C.}~\bibnamefont
  {Schneider}}, \bibinfo {author} {\bibfnamefont {J.}~\bibnamefont {Oertel}}, \
  and\ \bibinfo {author} {\bibfnamefont {R.}~\bibnamefont {Sch{\"u}tzhold}},\
  }\href {\doibase 10.1007/JHEP06(2017)043} {\bibfield  {journal} {\bibinfo
  {journal} {JHEP}\ }\textbf {\bibinfo {volume} {06}},\ \bibinfo {pages} {043}
  (\bibinfo {year} {2017})},\ \Eprint {http://arxiv.org/abs/1703.09203}
  {arXiv:1703.09203 [hep-th]} \BibitemShut {NoStop}%
%\%CITATION = ARXIV:1703.09203;\%\%
\bibitem [{\citenamefont {Massacrier}(2016)}]{massacrier2016first}%
  \BibitemOpen
  \bibfield  {author} {\bibinfo {author} {\bibfnamefont {L.}~\bibnamefont
  {Massacrier}} (\bibinfo {collaboration} {LHCb}),\ }in\ \href
  {http://inspirehep.net/record/1487542/files/arXiv:1609.06477.pdf} {\emph
  {\bibinfo {booktitle} {{4th Large Hadron Collider Physics Conference (LHCP
  2016) Lund, Sweden, June 13-18, 2016}}}}\ (\bibinfo {year} {2016})\ \Eprint
  {http://arxiv.org/abs/1609.06477} {arXiv:1609.06477 [nucl-ex]} \BibitemShut
  {NoStop}%
%\%CITATION = ARXIV:1609.06477;\%\%
\bibitem [{\citenamefont {Wilde}(2013)}]{alice2013measurement}%
  \BibitemOpen
  \bibfield  {author} {\bibinfo {author} {\bibfnamefont {M.}~\bibnamefont
  {Wilde}} (\bibinfo {collaboration} {ALICE}),\ }\bibfield  {booktitle} {\emph
  {\bibinfo {booktitle} {{Proceedings, 23rd International Conference on
  Ultrarelativistic Nucleus-Nucleus Collisions : Quark Matter 2012 (QM 2012):
  Washington, DC, USA, August 13-18, 2012}}},\ }\href {\doibase
  10.1016/j.nuclphysa.2013.02.079} {\bibfield  {journal} {\bibinfo  {journal}
  {Nucl. Phys.}\ }\textbf {\bibinfo {volume} {A904-905}},\ \bibinfo {pages}
  {573c} (\bibinfo {year} {2013})},\ \Eprint {http://arxiv.org/abs/1210.5958}
  {arXiv:1210.5958 [hep-ex]} \BibitemShut {NoStop}%
%\%CITATION = ARXIV:1210.5958;\%\%
\bibitem [{\citenamefont {Abelev}\ \emph {et~al.}(2013)\citenamefont {Abelev}
  \emph {et~al.}}]{abelev2013centrality}%
  \BibitemOpen
  \bibfield  {author} {\bibinfo {author} {\bibfnamefont {B.}~\bibnamefont
  {Abelev}} \emph {et~al.} (\bibinfo {collaboration} {ALICE}),\ }\href
  {\doibase 10.1103/PhysRevC.88.044909} {\bibfield  {journal} {\bibinfo
  {journal} {Phys. Rev.}\ }\textbf {\bibinfo {volume} {C88}},\ \bibinfo {pages}
  {044909} (\bibinfo {year} {2013})},\ \Eprint {http://arxiv.org/abs/1301.4361}
  {arXiv:1301.4361 [nucl-ex]} \BibitemShut {NoStop}%
%\%CITATION = ARXIV:1301.4361;\%\%
\bibitem [{\citenamefont {Roberts}(1986)}]{roberts1986dirac}%
  \BibitemOpen
  \bibfield  {author} {\bibinfo {author} {\bibfnamefont {L.~E.}\ \bibnamefont
  {Roberts}},\ }\href {\doibase 10.1007/BF02724243} {\bibfield  {journal}
  {\bibinfo  {journal} {Nuovo Cim.}\ }\textbf {\bibinfo {volume} {A92}},\
  \bibinfo {pages} {247} (\bibinfo {year} {1986})}\BibitemShut {NoStop}%
%\%CITATION = NUCIA,A92,247;\%\%
\bibitem [{\citenamefont {Dobbins}\ and\ \citenamefont
  {Roberts}(1993)}]{dobbins1993updated}%
  \BibitemOpen
  \bibfield  {author} {\bibinfo {author} {\bibfnamefont {T.}~\bibnamefont
  {Dobbins}}\ and\ \bibinfo {author} {\bibfnamefont {L.~E.}\ \bibnamefont
  {Roberts}},\ }\href {\doibase 10.1007/BF02785619} {\bibfield  {journal}
  {\bibinfo  {journal} {Nuovo Cim.}\ }\textbf {\bibinfo {volume} {A106}},\
  \bibinfo {pages} {1295} (\bibinfo {year} {1993})}\BibitemShut {NoStop}%
%\%CITATION = NUCIA,A106,1295;\%\%
\bibitem [{\citenamefont {Reisenegger}(2003)}]{reisenegger2003origin}%
  \BibitemOpen
  \bibfield  {author} {\bibinfo {author} {\bibfnamefont {A.}~\bibnamefont
  {Reisenegger}},\ }in\ \href {http://www.if.ufrgs.br/hadrons/reisenegger1.pdf}
  {\emph {\bibinfo {booktitle} {{Proceedings, International Workshop on Strong
  Magnetic Fields and Neutron Star: Havanna, Cuba, April 6-13, 2003}}}}\
  (\bibinfo {year} {2003})\ pp.\ \bibinfo {pages} {33--49},\ \Eprint
  {http://arxiv.org/abs/astro-ph/0307133} {arXiv:astro-ph/0307133 [astro-ph]}
  \BibitemShut {NoStop}%
%\%CITATION = ASTRO-PH/0307133;\%\%
\bibitem [{\citenamefont {Pons}\ \emph {et~al.}(2009)\citenamefont {Pons},
  \citenamefont {Miralles},\ and\ \citenamefont
  {Geppert}}]{pons2008magnetothermal}%
  \BibitemOpen
  \bibfield  {author} {\bibinfo {author} {\bibfnamefont {J.~A.}\ \bibnamefont
  {Pons}}, \bibinfo {author} {\bibfnamefont {J.~A.}\ \bibnamefont {Miralles}},
  \ and\ \bibinfo {author} {\bibfnamefont {U.}~\bibnamefont {Geppert}},\ }\href
  {\doibase 10.1051/0004-6361:200811229} {\bibfield  {journal} {\bibinfo
  {journal} {Astron. Astrophys.}\ }\textbf {\bibinfo {volume} {496}},\ \bibinfo
  {pages} {207} (\bibinfo {year} {2009})},\ \Eprint
  {http://arxiv.org/abs/0812.3018} {arXiv:0812.3018 [astro-ph]} \BibitemShut
  {NoStop}%
%\%CITATION = ARXIV:0812.3018;\%\%
\bibitem [{\citenamefont {Potekhin}(2010)}]{potekhin2011physics}%
  \BibitemOpen
  \bibfield  {author} {\bibinfo {author} {\bibfnamefont {A.~Y.}\ \bibnamefont
  {Potekhin}},\ }\href {\doibase 10.3367/UFNe.0180.201012c.1279} {\bibfield
  {journal} {\bibinfo  {journal} {Phys. Usp.}\ }\textbf {\bibinfo {volume}
  {53}},\ \bibinfo {pages} {1235} (\bibinfo {year} {2010})},\ \bibinfo {note}
  {[Usp. Fiz. Nauk180,1279(2010)]},\ \Eprint {http://arxiv.org/abs/1102.5735}
  {arXiv:1102.5735 [astro-ph.SR]} \BibitemShut {NoStop}%
%\%CITATION = ARXIV:1102.5735;\%\%
\bibitem [{\citenamefont {Harvey}\ \emph {et~al.}(1986)\citenamefont {Harvey},
  \citenamefont {Ruderman},\ and\ \citenamefont {Shaham}}]{harvey1985effects}%
  \BibitemOpen
  \bibfield  {author} {\bibinfo {author} {\bibfnamefont {J.~A.}\ \bibnamefont
  {Harvey}}, \bibinfo {author} {\bibfnamefont {M.~A.}\ \bibnamefont
  {Ruderman}}, \ and\ \bibinfo {author} {\bibfnamefont {J.}~\bibnamefont
  {Shaham}},\ }\href {\doibase 10.1103/PhysRevD.33.2084} {\bibfield  {journal}
  {\bibinfo  {journal} {Phys. Rev.}\ }\textbf {\bibinfo {volume} {D33}},\
  \bibinfo {pages} {2084} (\bibinfo {year} {1986})}\BibitemShut {NoStop}%
%\%CITATION = PHRVA,D33,2084;\%\%
\bibitem [{\citenamefont {Turolla}\ \emph {et~al.}(2015)\citenamefont
  {Turolla}, \citenamefont {Zane},\ and\ \citenamefont
  {Watts}}]{turolla2015magnetars}%
  \BibitemOpen
  \bibfield  {author} {\bibinfo {author} {\bibfnamefont {R.}~\bibnamefont
  {Turolla}}, \bibinfo {author} {\bibfnamefont {S.}~\bibnamefont {Zane}}, \
  and\ \bibinfo {author} {\bibfnamefont {A.}~\bibnamefont {Watts}},\ }\href
  {\doibase 10.1088/0034-4885} {\bibfield  {journal} {\bibinfo  {journal}
  {Rept. Prog. Phys.}\ }\textbf {\bibinfo {volume} {78}},\ \bibinfo {pages}
  {116901} (\bibinfo {year} {2015})},\ \Eprint
  {http://arxiv.org/abs/1507.02924} {arXiv:1507.02924 [astro-ph.HE]}
  \BibitemShut {NoStop}%
%\%CITATION = ARXIV:1507.02924;\%\%
\bibitem [{\citenamefont {Ahlen}(1978)}]{ahlen1978stopping}%
  \BibitemOpen
  \bibfield  {author} {\bibinfo {author} {\bibfnamefont {S.~P.}\ \bibnamefont
  {Ahlen}},\ }\href {\doibase 10.1103/PhysRevD.17.229} {\bibfield  {journal}
  {\bibinfo  {journal} {Phys. Rev.}\ }\textbf {\bibinfo {volume} {D17}},\
  \bibinfo {pages} {229} (\bibinfo {year} {1978})}\BibitemShut {NoStop}%
%\%CITATION = PHRVA,D17,229;\%\%
\bibitem [{\citenamefont {Ahlen}\ and\ \citenamefont
  {Kinoshita}(1982)}]{ahlen1982calculation}%
  \BibitemOpen
  \bibfield  {author} {\bibinfo {author} {\bibfnamefont {S.~p.}\ \bibnamefont
  {Ahlen}}\ and\ \bibinfo {author} {\bibfnamefont {K.}~\bibnamefont
  {Kinoshita}},\ }\href {\doibase 10.1103/PhysRevD.26.2347} {\bibfield
  {journal} {\bibinfo  {journal} {Phys. Rev.}\ }\textbf {\bibinfo {volume}
  {D26}},\ \bibinfo {pages} {2347} (\bibinfo {year} {1982})}\BibitemShut
  {NoStop}%
%\%CITATION = PHRVA,D26,2347;\%\%
\bibitem [{\citenamefont {Bracci}\ and\ \citenamefont
  {Fiorentini}(1983)}]{bracci1983binding}%
  \BibitemOpen
  \bibfield  {author} {\bibinfo {author} {\bibfnamefont {L.}~\bibnamefont
  {Bracci}}\ and\ \bibinfo {author} {\bibfnamefont {G.}~\bibnamefont
  {Fiorentini}},\ }\href {\doibase 10.1016/0370-2693(83)91559-9} {\bibfield
  {journal} {\bibinfo  {journal} {Phys. Lett.}\ }\textbf {\bibinfo {volume}
  {124B}},\ \bibinfo {pages} {493} (\bibinfo {year} {1983})}\BibitemShut
  {NoStop}%
%\%CITATION = PHLTA,124B,493;\%\%
\bibitem [{\citenamefont {Bracci}\ and\ \citenamefont
  {Fiorentini}(1984)}]{bracci1983interactions}%
  \BibitemOpen
  \bibfield  {author} {\bibinfo {author} {\bibfnamefont {L.}~\bibnamefont
  {Bracci}}\ and\ \bibinfo {author} {\bibfnamefont {G.}~\bibnamefont
  {Fiorentini}},\ }\href {\doibase 10.1016/0550-3213(84)90566-2} {\bibfield
  {journal} {\bibinfo  {journal} {Nucl. Phys.}\ }\textbf {\bibinfo {volume}
  {B232}},\ \bibinfo {pages} {236} (\bibinfo {year} {1984})}\BibitemShut
  {NoStop}%
%\%CITATION = NUPHA,B232,236;\%\%
\bibitem [{\citenamefont {Derkaoui}\ \emph {et~al.}(1998)\citenamefont
  {Derkaoui}, \citenamefont {Giacomelli}, \citenamefont {Lari}, \citenamefont
  {Margiotta}, \citenamefont {Ouchrif}, \citenamefont {Patrizii}, \citenamefont
  {Popa},\ and\ \citenamefont {Togo}}]{derkaoui1998energy}%
  \BibitemOpen
  \bibfield  {author} {\bibinfo {author} {\bibfnamefont {J.}~\bibnamefont
  {Derkaoui}}, \bibinfo {author} {\bibfnamefont {G.}~\bibnamefont
  {Giacomelli}}, \bibinfo {author} {\bibfnamefont {T.}~\bibnamefont {Lari}},
  \bibinfo {author} {\bibfnamefont {A.}~\bibnamefont {Margiotta}}, \bibinfo
  {author} {\bibfnamefont {M.}~\bibnamefont {Ouchrif}}, \bibinfo {author}
  {\bibfnamefont {L.}~\bibnamefont {Patrizii}}, \bibinfo {author}
  {\bibfnamefont {V.}~\bibnamefont {Popa}}, \ and\ \bibinfo {author}
  {\bibfnamefont {V.}~\bibnamefont {Togo}},\ }\href {\doibase
  10.1016/S0927-6505(98)00016-4} {\bibfield  {journal} {\bibinfo  {journal}
  {Astropart. Phys.}\ }\textbf {\bibinfo {volume} {9}},\ \bibinfo {pages} {173}
  (\bibinfo {year} {1998})}\BibitemShut {NoStop}%
%\%CITATION = APHYE,9,173;\%\%
\bibitem [{\citenamefont {Thompson}\ and\ \citenamefont
  {Duncan}(1993)}]{thompson1993neutron}%
  \BibitemOpen
  \bibfield  {author} {\bibinfo {author} {\bibfnamefont {C.}~\bibnamefont
  {Thompson}}\ and\ \bibinfo {author} {\bibfnamefont {R.~C.}\ \bibnamefont
  {Duncan}},\ }\href {\doibase 10.1086/172580} {\bibfield  {journal} {\bibinfo
  {journal} {Astrophys. J.}\ }\textbf {\bibinfo {volume} {408}},\ \bibinfo
  {pages} {194} (\bibinfo {year} {1993})}\BibitemShut {NoStop}%
%\%CITATION = ASJOA,408,194;\%\%
\bibitem [{\citenamefont {Hook}\ and\ \citenamefont
  {Huang}(2017)}]{hook2017bounding}%
  \BibitemOpen
  \bibfield  {author} {\bibinfo {author} {\bibfnamefont {A.}~\bibnamefont
  {Hook}}\ and\ \bibinfo {author} {\bibfnamefont {J.}~\bibnamefont {Huang}},\
  }\href@noop {} {\  (\bibinfo {year} {2017})},\ \Eprint
  {http://arxiv.org/abs/1705.01107} {arXiv:1705.01107 [hep-ph]} \BibitemShut
  {NoStop}%
%\%CITATION = ARXIV:1705.01107;\%\%
\end{thebibliography}%

\end{document}